# State of the Art in Direct Dark Matter Detectors: Technologies, Performance, and Future Directions

Muad Ghaith[a, b, *], Khaled Obaideen[c, d]

[a] Department of Physics, Queen's University, Kingston, ON K7L 3N6, Canada

[b] College of Natural and Health Sciences, Zayed University, Dubai 19282, U.A.E

[c] Research Institute of Sciences and Engineering, University of Sharjah, Sharjah, U.A.E

[d] College of Computing and Intelligent Systems, University of Al Dhaid, Al Dhaid, U.A.E

## Abstract

Identifying the particle nature of dark matter remains one of the most significant challenges in modern physics. Direct detection experiments aim to observe rare scattering events between dark matter particles and terrestrial targets, a task that demands extreme background suppression and sensitivity to minute energy depositions. This review critically assesses the current experimental landscape, organizing detector technologies by the fundamental physical trade-offs that define their scientific reach. We contrast the multi-tonne scalability of noble-liquid Time Projection Chambers (TPCs), which currently define the sensitivity frontier for high-mass Weakly Interacting Massive Particles (WIMPs), with the precision of cryogenic semiconductors and Charge-Coupled Device (CCD)- based sensors, which dominate the search for low-mass and sub-GeV candidates. Special emphasis is placed on the role of advanced reconstruction pipelines and machine learning (ML) as integral components of detector performance. Finally, we discuss the strategic roadmap for the next decade as experiments approach the neutrino fog, where coherent elastic neutrino-nucleus scattering ( CE$\nu$NS ) produces an increasingly significant and ultimately irreducible background that can mimic DM-induced nuclear recoils. We argue that future progress will rely not on a single technology but on a complementary global program combining increased target mass, ultra-low energy thresholds, improved background discrimination, and distinct observables - such as directionality and temporal signatures - to maintain robust discovery capability in the presence of neutrino-induced backgrounds.

*Corresponding author: muad.ghaith@queensu.ca

## 1. Introduction

### 1.1. *What is dark matter, and why do we search for it*

A broad range of astrophysical and cosmological observations consistently indicate that most matter in the Universe is non-luminous and interacts weakly, if at all, with electromagnetic radiation [1], [2]. This component, known as dark matter (DM), is required to explain the observed rotation curves of galaxies, gravitational lensing in galaxy clusters, anisotropies in the cosmic microwave background, and the formation of large-scale cosmic structure [3]. These phenomena are consistently described within a framework in which DM accounts for approximately 85% of the total matter content of the Universe [4].

While the gravitational influence of DM is firmly established, its fundamental particle nature remains unknown. Identifying the microscopic properties of DM, including its mass, interaction channels, and coupling strength to Standard Model particles, would constitute a major advance in fundamental physics[5]. Such a discovery would have profound implications for particle physics, cosmology, and our understanding of physics beyond the Standard Model [6].

### *1.2. Basic idea of direct detection*

Direct-detection experiments aim to observe DM through its rare interactions with ordinary matter in terrestrial detectors. A small fraction of these particles is expected to scatter elastically or inelastically off atomic nuclei or electrons in the detector target, depositing minute amounts of energy [7], [8].

The central challenge of direct detection stems from the combination of extremely low predicted interaction rates and the minute magnitude of the resulting energy depositions, typically at the keV scale or below [9]. Consequently, success requires a dual strategy: deploying ultra-sensitive detector technologies that can register these faint signals and suppressing environmental backgrounds by many orders of magnitude. This requirement drives the field toward deep-underground operations, rigorous material selection, and the development of sophisticated signal-background discrimination techniques.

### *1.3. Aim of this review*

This review offers a critical, systematic, and comparative assessment of the current state of direct DM detection. Rather than compiling an exhaustive catalog of experimental results, we focus on the detector technologies, signal channels, reconstruction methods, and performance-limiting factors that fundamentally determine experimental sensitivity.

We organize this analysis within a unified, technology-agnostic framework to enable direct comparison across a wide range of target materials and detector architectures. We use central performance criteria - including energy threshold, background discrimination, exposure, and reconstruction capability - to examine how sensitivity varies across DM mass and coupling regimes. This approach clarifies the physical trade-offs that shape detector design, identifies the domains where specific technologies are most effective, and highlights the complementarity among experimental strategies.

Crucially, we treat event reconstruction and data analysis, including the growing influence of advanced statistical and ML techniques, as integral to detector performance rather than as auxiliary processing tools. By systematically linking current experimental capabilities to future developments, such as detector scaling and ultra-low-threshold operation, this review distills general design principles that inform both the interpretation of existing results and the development of next-generation direct DM searches.

### 1.4. Structure of the paper

The review is organized as follows. Section 2 introduces the fundamental principles of direct DM detection and defines the key performance metrics that determine experimental sensitivity. Section 3 reviews the major detector technologies currently deployed or under active development, emphasizing their operating principles, strengths, and limitations, together with the underground infrastructure, shielding, and background mitigation. Section 4 discusses event reconstruction, calibration, and data analysis methods across detector classes. Section 5 presents a comparative assessment of experimental sensitivities, background rejection capabilities, and physics reach across interaction channels and mass regimes. Finally, Section 6 outlines future directions in detector development and discusses the outlook for next-generation direct DM searches.

## 2. Principles of Direct Detection

Direct DM detection experiments aim to measure the small amount of energy deposited when a DM particle scatters off ordinary matter [10]. Although interactions with Standard Model particles are predicted to be extremely weak, rare elastic or inelastic scattering off atomic nuclei or electrons may occur in sufficiently sensitive targets [11].

When an interaction occurs, a small amount of energy, typically on the keV scale or lower, is deposited in the target medium [12], [13]. The observables from a given interaction depend on the detector technology, which determines achievable energy thresholds, background-rejection capabilities, and event-reconstruction fidelity [14], [15].

Because expected interaction rates are often well below one event per kilogram per year, experiments must operate in environments with exceptionally low background radiation [16]. This requires deep underground installation to suppress cosmic-ray–induced backgrounds, rigorous selection of radiopure materials, and the use of passive shielding and active veto systems [17], [18].

### 2.1. The Basic Detection Principle (Light, Charge, Heat)

Direct detection relies on identifying how deposited energy manifests in a detector medium. The dominant signal channels are scintillation light, ionization charge, phonons (heat), and spatially extended recoil tracks.

Measuring these channels, either individually or in combination, defines the detector's scientific reach as shown in Figure 1. Experiments that measure two simultaneous signal channels generally achieve the strongest background rejection. By comparing the ratios of energy partitioned among different channels (e.g., ionization vs. phonons or scintillation vs. ionization), these detectors can statistically distinguish nuclear recoils (NR) (the expected WIMP signature) from the dominant electron-recoil (ER) background [19], [20].

Different detector classes optimize these channels for distinct regions of the search parameter space:

- Dual-channel detectors (e.g., Noble Liquid TPCs and Cryogenic

Bolometers) prioritize background discrimination and position reconstruction [21].

- Threshold-optimized detectors (e.g., CCDs and Point-Contact Germanium) are designed to resolve minute ionization or phonon signals to probe low-mass DM, often at the expense of discrimination power [22], [23].
- Topology-sensitive detectors (e.g., Directional Gas TPCs, Bubble Chambers) exploit the spatial structure or thermodynamic density of the energy deposit to reject backgrounds based on event class rather than spectral ratio [24].
- Modulation-sensitive detectors (e.g., NaI(Tl) scintillators such as DAMA/LIBRA, ANAIS, and COSINE) emphasize long-term stability, low, well-characterized backgrounds, and stable energy calibration to search for the annual modulation expected from Earth's motion through the Galactic DM halo [25] [26].

### *2.2. Key Performance Concepts*

Three central concepts define the sensitivity of a direct detection experiment:

1. Energy Threshold: The minimum energy deposition that can be reliably distinguished from detector noise. Lower thresholds are critical for sensitivity to low-mass DM particles (<10 GeV), which produce correspondingly small recoil energies [27].
2. Exposure: Defined as the product of target mass and live time. For rare-event searches, sensitivity generally scales linearly with exposure when backgrounds are negligible (the "zero-background" regime).
3. Background rejection and Position Reconstruction: The ability to suppress non-DM interactions, such as radioactive decays and cosmogenic interference, while accurately localizing interactions within the detector volume. Position reconstruction enables fiducialization, i.e., defining a cleaner inner volume effectively shielded by the detector's outer layers, which further enhances background rejection by excluding events near detector boundaries [28], [29].

### *2.3. The Role of Reconstruction and Analysis*

Reconstruction algorithms translate raw sensor outputs, such as voltage pulses, photon counts, or pixel clusters, into physical observables. The complexity of these pipelines scales with the readout architecture (Table 1).

In multi-channel detectors, reconstruction is not merely data processing but a primary discrimination tool [30]. For example, noble-liquid TPCs use the time delay between primary scintillation and secondary ionization signals to reconstruct the $z$-coordinate of an event, while the hit pattern on the photosensors provides the $xy$ position [31]. Similarly, cryogenic detectors use phonon pulse-shape timing to identify surface events that might mimic bulk NR [21].

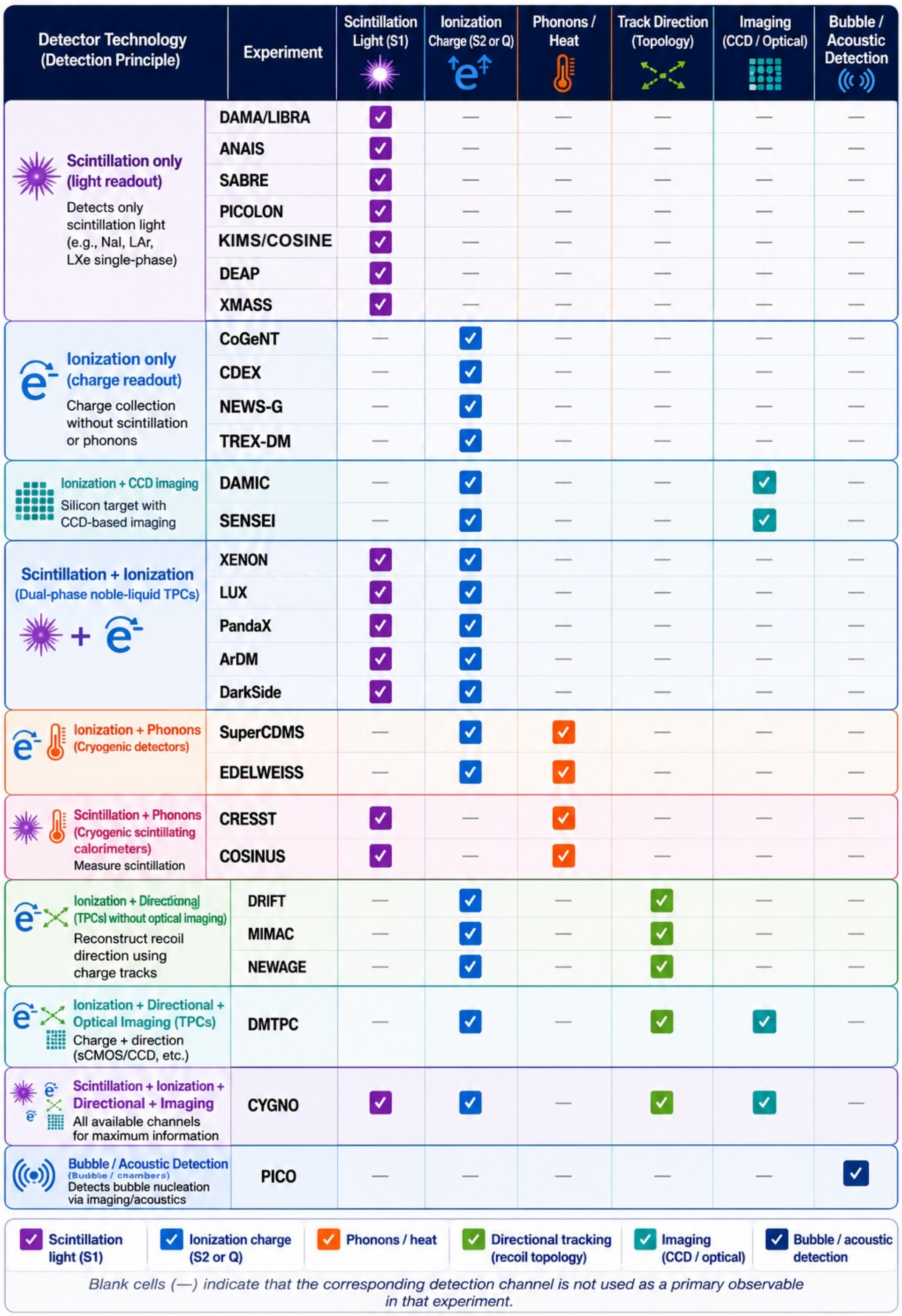

| Detector Technology (Detection Principle) | Experiment | Scintillation Light (S1) | Ionization Charge (S2 or Q) | Phonons / Heat | Track Direction (Topology) | Imaging (CCD / Optical) | Bubble / Acoustic Detection |
|---|---|---|---|---|---|---|---|
| Scintillation only (light readout) Detects only scintillation light (e.g., NaI, LAr, LXe single-phase) | DAMA/LIBRA | ✓ | — | — | — | — | — |
| | ANAIS | ✓ | — | — | — | — | — |
| | SABRE | ✓ | — | — | — | — | — |
| | PICOLON | ✓ | — | — | — | — | — |
| | KIMS/COSINE | ✓ | — | — | — | — | — |
| | DEAP | ✓ | — | — | — | — | — |
| | XMASS | ✓ | — | — | — | — | — |
| Ionization only (charge readout) Charge collection without scintillation or phonons | CoGeNT | — | ✓ | — | — | — | — |
| | CDEX | — | ✓ | — | — | — | — |
| | NEWS-G | — | ✓ | — | — | — | — |
| | TREX-DM | — | ✓ | — | — | — | — |
| Ionization + CCD imaging Silicon target with CCD-based imaging | DAMIC | — | ✓ | — | — | ✓ | — |
| | SENSEI | — | ✓ | — | — | ✓ | — |
| Scintillation + Ionization (Dual-phase noble-liquid TPCs) | XENON | ✓ | ✓ | — | — | — | — |
| | LUX | ✓ | ✓ | — | — | — | — |
| | PandaX | ✓ | ✓ | — | — | — | — |
| | ArDM | ✓ | ✓ | — | — | — | — |
| | DarkSide | ✓ | ✓ | — | — | — | — |
| Ionization + Phonons (Cryogenic detectors) | SuperCDMS | — | ✓ | ✓ | — | — | — |
| | EDELWEISS | — | ✓ | ✓ | — | — | — |
| Scintillation + Phonons (Cryogenic scintillating calorimeters) Measure scintillation | CRESST | ✓ | — | ✓ | — | — | — |
| | COSINUS | ✓ | — | ✓ | — | — | — |
| Ionization + Directional (TPCs) without optical imaging) Reconstruct recoil direction using charge tracks | DRIFT | — | ✓ | — | ✓ | — | — |
| | MIMAC | — | ✓ | — | ✓ | — | — |
| | NEWAGE | — | ✓ | — | ✓ | — | — |
| Ionization + Directional + Optical Imaging (TPCs) Charge + direction (sCMOS/CCD, etc.) | DMTPC | — | ✓ | — | ✓ | ✓ | — |
| Scintillation + Ionization + Directional + Imaging All available channels for maximum information | CYGNO | ✓ | ✓ | — | ✓ | ✓ | — |
| Bubble / Acoustic Detection (Bubble / chambers) Detects bubble nucleation via imaging/acoustics | PICO | — | — | — | — | — | ✓ |

Figure 1: Overview of detection channels used by major direct dark matter experiments. Experiments are classified according to their primary measured signals. Image resources: [32]–[56].

In single-channel or threshold-focused detectors, analysis emphasizes event classification rather than precise energy reconstruction. Bubble chambers, for instance, use acoustic parameters to distinguish alpha decays from NR, while CCD-based experiments rely on topological cluster analysis to identify low-energy electron tracks.

Table 1. Data acquisition (DAQ) methods and readout characteristics of representative detectors, highlighting sampling, triggering, and event-rate behavior.

| Experiment | Sampling / Readout Style | Event Rate Features | Ref. |
|---|---|---|---|
| ANAIS | Continuous Photomultiplier Tube (PMT) waveform digitization with coincidence triggering | Low event rate; annual modulation search requiring long-term stability | [32] |
| COSINE | Continuous PMT waveform digitization of NaI(Tl) crystals with coincidence-based triggering and liquid-scintillator veto readout | Low event rate; annual modulation search with stringent long-term stability and background rejection | [57] |
| COSINUS | Simultaneous phonon and scintillation readout using TES | Very low event rate; event-by-event particle discrimination | [36] |
| CRESST | Continuous TES pulse acquisition with triggered readout | Extremely low event rate; optimized for low-energy NR | [37] |
| CYGNO | High-speed optical imaging synchronized with PMT signals | Low event rate; directional track reconstruction | [38] |
| DAMIC | CCD exposures read out periodically | Low event rate, dominated by backgrounds | [40] |
| DEAP | Scintillation pulses with fine timing resolution | Relies on detailed Pulse-Shape Discrimination (PSD) over many photoelectrons | [58] |
| LAr TPCs (DarkSide/ArDM) | Dual-phase readout combining PMT array | High-rate handling capability; robust background rejection via light-to-charge ratios | [33], [59], [60] |
| LXe TPCs (XENON/LUX/ PandaX) | PMT waveforms digitized at tens–hundreds of MHz | Triggers on S1/S2; continuous low-rate operation | [61] |
| NEWAGE | Micro-pattern TPC readout with time rejection | Low event rate; directional recoil reconstruction | [46] |
| NEWS-G | Charge pulses from spherical proportional counter | Very low event rate; optimized for light DM | [47] |
| PICO | Camera frames and acoustic sensors on bubble nucleation | Low duty cycle for event triggers | [48] |
| PICOLON | PMT waveform acquisition with coincidence triggering | Low event rate; annual modulation analysis over long exposure | [54] |
| SABRE | PMT waveform digitization with active-veto coincidence | Low event rate; long-term annual modulation | [49] |
| SENSEI | MHz sampling of phonon channels with triggered readout | Very low DM event rate; background dominated | [50] |
| SuperCDMS | MHz sampling of phonon channels with triggered readout | Very low DM event rate; background-dominated | [62] |
| TREX-DM | Continuous micromegas waveform digitization | Low event rate; optimized for low-mass DM | [55] |

### 2.4. *State of the Art by Detector Technology*

The field has not converged on a single experimental design; instead, a portfolio of complementary technologies has been developed to cover the wide range of potential DM masses and coupling strengths [63].

The following sections review the dominant detector classes:

- Cryogenic Semiconductors: operated at millikelvin temperatures to achieve ultra-low energy thresholds, providing excellent sensitivity to low-mass DM candidates through the measurement of phonons, ionization, or both [64].
- Pure Ionization Detectors: Semiconductor detectors that rely exclusively on the collection of ionization charge, enabling precise energy measurements and low-threshold searches for light DM, particularly through ER signatures [65].

- Bubble Chambers: Superheated liquid detectors that offer exceptional sensitivity to spin-dependent (SD) interactions through their intrinsic rejection of electron recoil (ER) backgrounds [66].
- Noble-Liquid TPCs: Large-scale detectors employing liquid xenon or argon to simultaneously measure scintillation and ionization, delivering the most stringent constraints on high-mass spin-independent (SI) DM interactions [67].
- Scintillating Crystals: Radiopure solid-state detectors that measure scintillation light with excellent long-term stability, making them well suited for time-dependent searches such as annual modulation studies [68].
- CCD/Imaging Detectors: Silicon-based imaging technologies with extremely low noise and energy thresholds, offering leading sensitivity to sub-GeV DM through electron-scattering interactions [65].
- Gaseous Detectors: Gas-filled detectors operated at atmospheric or elevated pressures to measure ionization produced by particle interactions. Their low target density and flexible readout technologies make them attractive for low-energy event detection and specialized DM searches, including light DM and rare interaction channels [69].
    - Directional Gas Detectors: Low-pressure gaseous detectors capable of reconstructing NR tracks, providing directional information that can distinguish a DM signal from isotropic backgrounds [70].

This technological diversity ensures that limitations inherent to one approach, such as the high gas-detector threshold or the scalability limits of bolometers, are offset by the strengths of other approaches.

## 3. Detector Technologies and Background Mitigation

Direct detection relies on a diverse portfolio of technologies, each optimized for specific regions of the DM parameter space (Table 2). Design choices for the target medium and readout architecture determine the achievable energy threshold, discrimination power, and background level [9]. Because the expected DM interaction rate is extremely low, these detectors typically operate in deep-underground laboratories and are surrounded by passive and/or active shielding to suppress cosmic-ray-induced and environmental radioactive backgrounds.

### *3.1. Cryogenic Semiconductor Detectors*

Cryogenic detectors use Germanium or Silicon crystals operated at millikelvin temperatures. In this regime, even minute energy depositions produce measurable temperature rises (phonons) alongside ionization charge [71].

By instrumenting these crystals with Transition Edge Sensors (TES) or similar superconducting thermometers, experiments such as SuperCDMS and EDELWEISS achieve excellent energy resolution. The primary engineering advantage of this architecture is the ability to distinguish surface events (often caused by ambient beta radiation) from bulk events through phonon pulse-shape analysis and timing. Although currently limited in total mass compared to noble liquids, these detectors can reject backgrounds near the energy threshold, making them the premier technology for low-mass (<10 GeV) WIMP searches [72].

### *3.2. Pure Ionization Detectors*

Pure ionization detectors measure the charge carriers produced by particle interactions without requiring a complementary phonon or scintillation signal. Semiconductor implementations based primarily on high-purity germanium (HPGe), including point-contact architectures used by experiments such as CoGeNT and CDEX, exploit their small readout capacitance to achieve low electronic noise and sub-keV ionization thresholds [73]–[75].

The principal advantage of this architecture is its combination of low threshold, excellent energy resolution, and comparatively simple readout. However, because recoil energy is inferred predominantly from ionization, these detectors generally lack the event-by-event electron-recoil/nuclear-recoil discrimination available in dual-channel phonon–ionization or scintillation–ionization detectors. Background control therefore relies heavily on material radiopurity, passive and active shielding, event topology or pulse-shape information, and detailed spectral background modeling. This trade-off makes pure ionization detectors particularly effective for low-mass DM searches, where achieving the lowest possible threshold can be more important than maintaining strong recoil discrimination [76].

### *3.3. Bubble Chambers*

Bubble chambers, such as PICO, employ superheated liquids (e.g., $C_3F_8$) maintained in a metastable state. When a sufficiently high localized energy density, such as that produced by an NR, is deposited, it induces a phase transition that forms a macroscopic bubble [77].

The thermodynamic parameters of the chamber are tuned so that ER (which deposit energy along longer tracks) do not trigger nucleation. This renders the instrument effectively blind to the dominant gamma and beta backgrounds. The observables are binary (bubble vs. no-bubble) and acoustic; piezoelectric sensors record the sound of bubble formation to discriminate between alpha decays and NR. This technology leads the field in SD proton sensitivity due to the high fluorine content of the target fluids [78]–[80].

### *3.4. Noble-Liquid Detectors*

Noble-liquid TPCs using liquid xenon (LXe) or argon (LAr) are the most scalable technology in the field. These cylindrical detectors apply a vertical electric field across a large volume of cryogenic liquid.

Interaction events produce prompt scintillation light ($S1$) and liberate ionization electrons. Under an applied electric field, the electrons drift toward a gas extraction region, where they generate a secondary electroluminescence signal ($S2$). This dual-signal readout enables full three-dimensional position reconstruction, thereby defining a low-background fiducial volume deep within the detector [67].

- Liquid xenon (e.g., XENON, LZ, PandaX): Provides high density ( $\sim 3\ \mathrm{g/cm^3}$ ) and excellent radiopurity and currently dominates sensitivity to high-mass SI interactions [81].
- Liquid Argon (e.g., DarkSide, DEAP): Enables powerful PSD based on scintillation timing, achieving extremely strong background rejection, though with a lower intrinsic atomic mass than XENON [82].

### *3.5. Scintillating Crystals*

Experiments such as DAMA/LIBRA, ANAIS, and COSINE use arrays of high-purity NaI(Tl) scintillating crystals read out by photomultiplier tubes, while earlier experiments such as KIMS employed CsI(Tl). Unlike TPCs or bolometers, these detectors typically rely on a single signal channel (scintillation), providing limited event-by-event background discrimination. Their primary strength is long-term stability and simplicity, making them ideal for testing the annual modulation hypothesis, the periodic fluctuation in event rate expected from Earth's orbit through the galactic halo [25], [83]–[86].

### *3.6. CCD / Imaging Detectors*

CCDs use the sensor's silicon bulk as the interaction target. Modern Skipper CCDs (e.g., DAMIC-M, SENSEI) employ non-destructive, repetitive readout techniques to suppress electronic noise to sub-electron levels. This capability enables detection of single-electron ionization events, effectively reducing the energy threshold to the silicon bandgap ($\sim 1.1\ \mathrm{eV}$). As a result, this detector architecture is uniquely suited to probe sub-GeV DM via electron scattering, a parameter space largely inaccessible to conventional NR detectors [87], [88].

### *3.7. Directional Gas Detectors*

Directional detectors (e.g., DRIFT, MIMAC, CYGNO) operate on gaseous targets at low pressure (<100 Torr). In this rarefied medium, NR travel sufficiently far (on the order of millimeters) to produce reconstructable ionization tracks.

By projecting the ionization charge onto a fine-grained readout plane (or using optical imaging), these detectors can reconstruct the recoil axis and "head-tail" orientation. This enables correlation with the direction of the Solar System's motion toward the Cygnus constellation. The trade-off is density: achieving tonne-scale mass with low-pressure gas requires immense detector volumes, which currently limit this technology to discovery confirmation rather than to exclusion limits [70].

Table 2. Main detector technologies, target materials, and core detection principles in direct dark matter searches.

| Experiment | Detector class | Target material | Core concept | Main readout | Ref. |
|---|---|---|---|---|---|
| ANAIS | Scintillating crystals | NaI(Tl) | Annual modulation | PMTs | [32] |
| ArDM | Dual-phase LAr TPC | Ar | LAr R&D for DM | PMTs + LEMs | [33] |
| CDEX | Semiconductor detector | Ge | Low-mass WIMP, sub-keV threshold | Point contact electrodes + preamps | [34], [89] |
| CoGeNT | HPGe diode | Ge | Low-noise ionization only | Point contact electrodes + preamps | [35], [73] |
| COSINE | Scintillating crystals | NaI(Tl) | Annual modulation; test of DAMA/LIBRA | PMTs + liquid-scintillator veto | [90] |
| COSINUS | Cryogenic scintillating calorimeters | NaI | Simultaneous phonon & scintillation | TES + light detector | [36] |
| CRESST | Cryogenic scintillating calorimeters | $CaWO_4$ | NR measured via phonon & scintillation | TES + light detector | [37] |
| CYGNO | Optical gas TPC | He/$CF_4$ | Directional gaseous detector + optical imaging | sCMOS Cameras + PMTs | [38] |
| DAMA/ LIBRA | Scintillating crystals | NaI(Tl) | Annual modulation | PMTs | [39], [91] |
| DAMIC | CCD | Si | Single-electron sensitivity | Pixel charge imaging | [92] |
| DarkSide | Dual-phase LAr TPC | Ar | S1/S2 + PSD | Silicon Photomultipliers (SiPMs) | [93] |
| DEAP | Single-phase LAr | Ar | PSD in LAr | High efficiency PMTs | [94] |
| DMTPC | Directional gas TPC | $CF_4$ | Low-pressure TPC; optical track imaging | CCD + charge readout | [41] |
| DRIFT | Directional gas TPC | $CS_2/CF_4$ | Negative-ion TPC; long tracks | Charge + time projection | [95] |
| EDELWEISS | Cryogenic bolometers | Ge | Heat + ionization with interleaved electrodes | NTD thermistors + electrodes | [96] |
| KIMS | Scintillating crystals | CsI(Tl) | Annual modulation | PMTs | [97] |
| LUX | Dual-phase LXe TPC | Xe | S1+S2 in LXe | PMTs | [81], [98] |
| MIMAC | Micro-TPC | $He/CF_4$ mixes | Directional detection, SD coupling | Micromegas readout | [45] |
| NEWAGE | Micro-pattern gas TPC | $CF_4$ | Direction-sensitive gaseous detectors | μ-PIC + TPC readout | [46] |
| NEWS-G | Spherical proportional counter | Ne, $CH_4$, He mixtures | Ultra-low-capacitance spherical gas detector | Central anode proportional readout | [99] |
| PandaX | Dual-phase LXe TPC | Xe | Multi-tonne LXe for SI sensitivity | Cryogenic PMTs | [100] |
| PICO | Bubble chamber | $C_3F_8$ | Superheated liquid; nucleation from NR | Acoustic sensors + optical cameras | [48] |
| PICOLON | Scintillating crystals | NaI(Tl) | High-purity NaI(Tl) annual modulation | PMTs | [54] |
| SABRE | Scintillating crystals | NaI(Tl) | Independent annual modulation | PMTs with active liquid scintillator veto | [49] |
| SENSEI | Skipper CCD | Si | Single-electron detection | Skipper CCD pixel readout | [50] |
| SuperCDMS | Cryogenic bolometers | Ge, Si | DM recoils as phonons + ionization | TES + charge electrodes | [101] |
| TREX-DM | High-pressure | Ar or Ne | Low-mass DM search with gaseous TPC | Micromegas readout | [55] |
| | micromegas TPC | | | | |
| XENON | Dual-phase LXe TPC | Xe | S1+S2 in LXe | PMTs, TPC | [102] |
| XMASS | Single-phase LXe | Xe | Scintillation only | PMTs on sphere | [103] |

### *3.8. Underground Environments and Shielding*

To achieve the required sensitivity, experiments must isolate the target from environmental radiation. This is accomplished through a layered defense strategy: depth, passive shielding, and active vetoes [28], [29].

Depth: Experiments are located deep underground to attenuate cosmic-ray muons. The shielding provided by the surrounding rock, or overburden, is quantified in meters of water equivalent (m.w.e.). As shown in Figure 2, experimental depths range from approximately 2000 m.w.e. (Soudan Underground Laboratory) to more than 6000 m.w.e. (SNOLAB; China Jinping Underground Laboratory), reducing the cosmic-muon flux by up to seven orders of magnitude [104].

Shielding & Vetoes:

- Passive Shielding: Layers of lead (for gamma rays) and polyethylene (for neutrons) surround the detector.
- Active Vetoes: Large detectors (especially noble-liquid experiments) often employ instrumented water tanks or dedicated scintillator systems as active vetoes. Water can provide passive shielding while surrounding PMTs detect Cherenkov radiation from passing muons, allowing the data acquisition system to "veto" (discard) any internal events coincident with muon passage [105]. Plastic or liquid scintillators can also be used as active veto systems, providing an additional layer of muon and external-background tagging around the detector.

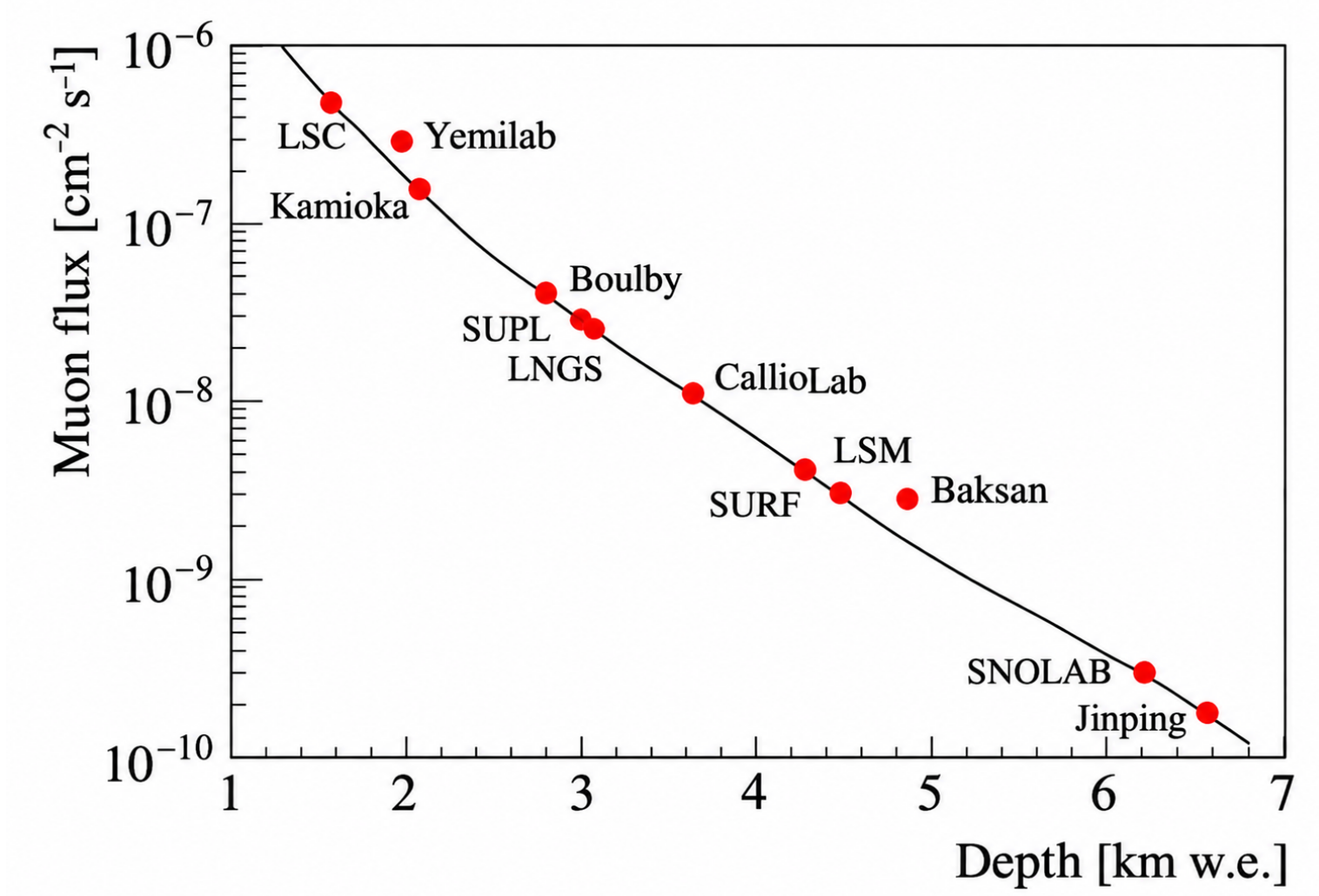

Figure 2. Muon flux measured in underground laboratories. The solid black curve shows the expected muon flux as a function of depth for a uniform (flat) overburden. In contrast, laboratories beneath mountainous terrain experience direction-dependent muon fluxes because rock thickness varies. For these sites, the figure above shows the maximum overburden, which exceeds the equivalent depth under flat-ground conditions. All laboratory data points are plotted against the solid reference curve for comparison. Image modified from source [104].

## 4. Reconstruction and Analysis

Reconstruction algorithms transform raw sensor signals into physical observables: deposited energy, interaction position, and particle classification. As detector hardware approaches fundamental noise limits, sensitivity improvements are increasingly driven by software: sophisticated filtering, likelihood inference, and ML pipelines that extract signal from noise with higher fidelity than traditional pulse-integral methods [106].

### *4.1. Baseline Signal Processing*

Despite technological differences, most analyses share a common foundation of signal-processing steps to ensure data quality [106].

Energy Estimation: In the high-signal-to-noise regime, energy is commonly derived from pulse integrals. However, linearity corrections are essential to account for saturation effects or non-linear detector responses at low energies.

Fiducialization: Surface backgrounds often dominate interaction rates. Algorithms define a "virtual detector" within the physical volume by reconstructing event positions and rejecting interactions near the walls.

Quality Cuts: Timing-based selection criteria suppress non-physical triggers, such as electronic microphonics or pile-up (two events occurring within the same integration window).

### *4.2. Reconstruction Architectures, Advanced Interference, and Machine Learning*

The complexity of event reconstruction scales with the richness and dimensionality of detector readout (Table 3). Modern rare-event searches increasingly move beyond conventional cut-and-count approaches toward information-rich reconstruction, multivariant inference, and ML techniques. These approaches can be broadly divided into information-dense inference and feature-based classification, although modern experiments increasingly combine both. Pipelines generally fall into one of two categories:

1. Information-Dense Inference: Multi-channel detectors, such as cryogenic bolometers and noble-liquid TPCs, exploit redundant observables to perform joint reconstruction and inference. Near the energy threshold, conventional pulse integration is often limited by electronic and detector noise. Experiments such as SuperCDMS and EDELWEISS therefore use optimal filters and template-fitting techniques, weighting frequency components by the inverse power spectrum. This maximizes the signal-to-noise ratio for a known pulse shape and enables near-minimum-variance energy and position estimates, extending sensitivity toward lower analysis thresholds. In noble-liquid TPCs, reconstructing scintillation (S1) and ionization (S2) signals combined provides additional spatial and particle-identification information [14]. The relative timing of S1 and S2 determines interaction depth, while the spatial distribution of the S2 signal provides transverse-position information. These observables are subsequently used to correct for position-dependent detector response and define fiducial volumes.
2. Feature-Based Classification and Multivariate Inference: Detectors with more limited spectroscopic information, such as bubble chambers and imaging detectors, often rely more strongly on topological or morphological features. Bubble-chamber analysis can use acoustic and event-topology information to discriminate nuclear recoils from background events, whereas CCD-based experiments exploit pixel-level charge patterns, cluster morphology, and spatial distributions. More generally, even when signal and background populations overlap in a single observable, such as the S2/S1 ratio in noble-liquid detectors, they may remain separable in higher-dimensional feature space. Multivariate classifiers, including boosted decision trees, reconstruct quality and other event observables into a single discriminating variable. These techniques complement profile-likelihood approaches, which incorporate multiple observables directly into a statistical inference framework [106].
3. Machine Learning and Validation: ML methods provide increasingly powerful tools for event reconstruction, background discrimination, and identifying non-physical detector artifacts. For example, deep-learning approaches can classify complex noise patterns and charge clusters in imaging detectors, while ML algorithms can improve position or energy reconstruction in multi-channel detectors. However, their application to rare-event searches requires particularly rigorous validation because small biases can have a significant impact when the expected signal rate is extremely low. Consequently, ML-based reconstruction and classification must be validated using independent calibration data, control samples, and physically motivated detector simulations. This ensures that the algorithm improves discrimination without introducing artificial signal-like structures or biasing the reconstructed event population [106], [107].

Overall, modern reconstruction pipelines increasingly combine signal-processing techniques for precise low-energy reconstruction, multivariate methods for background discrimination, and ML for complex, high-dimensional event classification. The appropriate architecture depends strongly on the detector technology, available observables, and dominant background populations.

Table 3. Examples of event reconstruction and analysis methods across direct DM search experiments.

| Experiment | Event reconstruction and analysis method | Ref. |
|---|---|---|
| ANAIS | Pulse-area and photoelectron extraction, energy calibration, event-quality selection, and time-series analysis for annual modulation | [32], [108], [109] |
| COSINE | PMT waveform/photoelectron reconstruction, energy calibration, pulse-shape discrimination, BDT-based noise rejection, event selection, and time-series modulation analysis | [110] |
| COSINUS | Combined phonon-light signal reconstruction, pulse-shape discrimination, likelihood analysis | [36] |
| CRESST | Optimal filtering of TES pulses, simultaneous phonon-light reconstruction, likelihood-based event classification | [111] |
| CYGNO | Optical image reconstruction, CNN-based track classification, charge/light correlation analysis | [38] |
| DAMA /LIBRA | Pulse-area/photoelectron extraction, spectra and time-series modulation analysis | [112] |
| DAMIC | Image reconstruction, pixel-cluster finding and topology classification | [113] |
| DarkSide | S1/S2 pulse reconstruction, pulse-area and pulse-time extraction, position reconstruction from S2 light pattern, pulse-shape discrimination, and S1/S2-based event classification | [114] |
| DEAP | Photon/pulse reconstruction, position reconstruction using PMT charge and photon timing, and fiducial volume selection | [115], [116] |
| DRIFT | 3D negative-ion TPC track reconstruction using charge arrival times and segmented readout, with recoil-track direction and topology reconstruction | [117] |
| DMTPC | Optical image reconstruction, charge/ amplitude extraction, 2D track and head-tail reconstruction | [118], [119] |
| EDELWEISS | Pulse fitting and likelihood analysis in the heat-ionization plane | [96] |
| LUX | S1/S2 pulse reconstruction, S2-based $x-y$ position reconstruction, S1-S2 timing for $z$ reconstruction, pulse-quality and event-selection criteria | [120] |
| MIMAC | 3D track reconstruction from time-sampled strips, recoil-energy reconstruction, and directional/angular analysis | [121], [122] |
| NEWAGE | μ-PIC track reconstruction, direction fitting, angular analysis | [123], [124] |
| NEWS-G | Pulse-shape analysis, charge spectrum reconstruction, likelihood-based event selection | [125] |
| PandaX | S1/S2 signal reconstruction, drift-time and S2 pattern-based 3D position reconstruction, signal/background discrimination | [126] |
| PICO | Waveform and acoustic-parameter analysis | [48] |
| PICOLON | Pulse-area reconstruction, energy calibration, annual modulation analysis | [54] |
| SABRE | PMT pulse reconstruction, event selection, long-term stability monitoring | [127] |
| SENSEI | Pixel-level charge counting, cluster identification, single-electron reconstruction | [50] |
| SuperCDMS | Optimal filters, multi-template fits, advanced multi-channel ML | [107] |
| XENON | S1/S2 pulse identification and reconstruction, position reconstruction from S1-S2 drift time and S2 light pattern, position-dependent signal corrections | [128]–[130] |
| XMASS | PMT charge reconstruction, position reconstruction, likelihood-based event selection | [131] |

### *4.3. Calibration Strategies*

Robust event reconstruction requires calibration to map detector response to physical energy scales. Calibration establishes the recoil-energy scale, characterizes detector response, and helps define signal and background regions [10], [132].

Electronic recoils (ER): These are typically calibrated using external gamma-ray sources (e.g., $^{60}Co$, $^{137}Cs$) or internal gaseous sources (e.g., $^{83m}Kr$, $^{220}Rn$). Internal calibration sources are especially important for large noble-liquid detectors because they provide a uniform activity distribution throughout the detector volume, enabling precise mapping of spatial variations in detector response [133].

Nuclear recoils (NR): These serve as a proxy for the WIMP signal. Experiments typically use neutron sources such as AmBe or deuterium-deuterium (DD) generators to populate the NR band [134], [135].

However, producing well-characterized low-energy NRs is technically challenging. Below the range of direct calibration measurements, analyses may rely on response models such as Lindhard theory extrapolated to lower energies, introducing additional systematic uncertainties [136], [137].

For experiments such as ANAIS, DAMA/LIBRA, KIMS, and SABRE, neutron calibrations are generally not part of the routine physics data analysis for event-by-event nuclear-recoil discrimination, since these NaI(Tl)-based experiments primarily search for an annual modulation signal. Nevertheless, dedicated neutron measurements characterize the nuclear-recoil response and determine quenching factors, which relate nuclear-recoil energies to the observed electron-equivalent energy scale, as listed in Table 4. Other experiments like COSINUS and CRESST employ cryogenic calorimeters, where phonon calibration (heater pulses) is as important as external radioactive sources. SENSEI, DAMIC, and NEWS-G emphasize low-energy electron-response calibration because of their low-threshold operation. Lastly, CYGNO, NEWAGE, and TREX-DM require additional calibration of gas gain and track reconstruction because of their gaseous TPC designs.

Table 4. Calibration approaches for electronic and NR, including sources and special calibration techniques.

| Experiment | ER Calibration sources | NR calibration sources | Special calibrations | Ref. |
|---|---|---|---|---|
| ANAIS | $^{109}Cd$, $^{57}Co$, internal $^{40}K$ & $^{22}Na$ | $^{252}Cf$ | Light-yield stability and gain monitoring | [138], [139] |
| COSINE | $^{241}Am$, $^{57}Co$, $^{60}Co$ & $^{137}Cs$ | Neutron-beam measurements on NaI(Tl) crystals for Na/I recoil quenching-factor characterization | Light-yield/gain stability monitoring; low-energy calibration using internal peaks; NaI(Tl) nonproportional-response calibration | [90], [140] |
| COSINUS | $^{55}Fe$, $^{57}Co$, & $^{241}Am$ | $^{241}AmBe$ | Simultaneous phonon-light calibration | [36] |
| CRESST | $^{57}Co$, $^{55}Fe$, $X$-ray sources, heater pulses | $^{241}AmBe$ and DD neutrons | Heater-pulse energy calibration | [111] |
| | | | and phonon/light yield calibration | |
| CYGNO | $^{55}Fe$, $X$-ray sources | $^{252}Cf$ | Optical gain calibration and track reconstruction | [141] |
| DAMA/LIBRA | $^{241}Am$, $^{57}Co$, $^{133}Ba$ & $^{137}Cs$ | No explicit NR calibration in data analysis | Long-term detector stability and energy-scale monitoring | [91] |
| DAMIC | $^{55}Fe$, $^{241}Am$ and cosmogenic $X$-ray lines | $^{252}Cf$ | Charge-transfer efficiency, single-electron response and spatial calibration | [142] |
| DarkSide | $^{37}Ar$, $^{39}Ar$, $^{22}Na$, & $^{57}Co/^{60}Co$ | $^{241}AmBe$ & $^{252}Cf$ | Underground $^{39}Ar$ characterization and PSD calibration | [93] |
| DEAP | $^{39}Ar$, $^{22}Na$ & $^{60}Co$ | $^{241}AmBe$ | Extensive PSD calibration | [143] |
| EDELWEISS | $^{60}Co$ & $^{133}Ba$ | $^{241}AmBe$ & $^{252}Cf$ | Surface-event studies using dedicated runs | [96] |
| KIMS | $^{241}Am$, $^{137}Cs$ | $^{241}AmBe$ | NR/ER pulse-shape discrimination; CsI recoil quenching-factor characterization | [97], [144] |
| LUX | $^{137}Cs$ / $^{60}Co$, $^{83m}Kr$, & tritium | $^{252}Cf$ & DD neutrons | $^{83m}Kr$ for spatial-response calibration & tritium ER calibration | [145] |
| NEWS-G | $^{37}Ar$, $^{55}Fe$ & $^{109}Cd$ | $^{241}AmBe$ | Gas-gain and spherical proportional-counter calibration | [146] |
| NEWAGE | $^{55}Fe$ & $^{109}Cd$ | $^{252}Cf$ | Directional response and gas-gain calibration | [46] |
| PandaX | $^{220}Rn$ / $^{222}Rn$ , $^{83m}Kr,$ & tritium | $^{241}AmBe$ & DD neutrons | $^{83m}Kr$ for spatial-response & electron-lifetime calibration | [147] |
| PICO | None (ERs strongly rejected by the bubble-nucleation thresholds) | $^{241}AmBe$ & $^{252}Cf$ | Acoustic response characterization | [48] |
| PICOLON | $^{109}Cd$, $^{60}Co$ & $^{133}Ba$ | $^{252}Cf$ | Light-yield and energy-scale calibration | [148] |
| SABRE | $^{57}Co$, $^{241}Am$, & $^{109}Cd$ | $^{252}Cf$ | Crystal background and stability calibration | [149] |
| SENSEI | $^{55}Fe$, $^{241}Am$ LED/photon injection | No dedicated NR calibration routinely used | Single-electron response and skipper-CCD calibration | [150] |
| SuperCDMS | $^{55}Fe$, $^{241}Am$, $^{57}Co$ & $^{133}Ba$ | $^{252}Cf$ & DD neutrons | Internal activation lines in Ge and detector-channel calibration | [72] |
| TREX-DM | $^{109}Cd$, $^{55}Fe$ & $X$-ray sources | $^{252}Cf$ | Gas-gain, Micromegas response, and threshold calibration | [151], [152] |
| XENON | $^{137}Cs$ / $^{60}Co$, $^{220}Rn$ / $^{222}Rn$ , $^{83m}Kr$, $^{37}Ar$, & tritium | $^{241}AmBe$ & DD neutrons | Internal $^{83m}Kr$ for spatial uniformity | [130] |

### *4.4. The Sensitivity Gain*

Advanced reconstruction directly affects an experiment's "Physics Reach." By stabilizing energy estimators, pipelines lower the effective threshold, thereby increasing sensitivity to lower-mass DM. By improving discrimination through multivariate analysis, they reduce the expected background count in the region of interest, allowing the experiment to run longer without background. Consequently, investment in analysis software often yields sensitivity improvements comparable to significant hardware upgrades [153].

## 5. Comparative Analysis Across All Experiments

Multiple factors determine the sensitivity of a direct-detection experiment,

including target mass and composition, energy threshold, exposure, and background level and discrimination. Consequently, no single detector technology provides optimal sensitivity across the full DM parameter space, motivating a complementary portfolio of experimental approaches [8], [154]. This section compares current experiments and maps the principal detector technologies to the DM mass regimes and interaction channels where they provide the greatest sensitivity.

### *5.1. The Interplay of Mass and Exposure*

For conventional WIMP scenarios, the expected interaction rates are extremely low and can fall well below those accessible in the parameter space probed by modern direct-detection experiments [155]. In the background-free regime, where the expected number of residual background events remains negligible, the cross-section sensitivity improves approximately inversely with (detector mass × live time), corresponding to a linear improvement in experimental reach with increasing exposure [8]. Once background events become significant, however, the sensitivity generally improves more slowly, approximately with the square root of exposure in the statistics-dominated regime. The relevant energy scale is often expressed in keV electron-equivalent ($keV_{ee}$), which denotes the amount of energy deposited by an electron recoil that produces the same measured detector response as the interaction under consideration. This convention is particularly useful when comparing detector thresholds and calibration based on electron-recoil sources.

Noble-liquid dominance: Liquid-xenon and liquid-argon detectors can achieve very large target masses and exposures (Table 5). Their fluid nature enables continuous purification and recirculation, while xenon's high liquid density (~3 $g/cm^3$) allows compact, multi-tonne target volumes [156], [157]. The large target volume, together with liquid xenon's high stopping power, provides effective self-shielding: external radioactive backgrounds are preferentially absorbed in the outer layers, allowing a low-background inner fiducial volume. Combined with scalability, three-dimensional position reconstruction, and strong ER/NR discrimination, these properties have enabled noble-liquid TPCs to achieve leading sensitivity to spin-independent WIMP interactions at higher DM masses [157], [158].

The Low-Mass Trade-off: Cryogenic semiconductor and CCD experiments typically operate with active detector masses ranging from grams to kilograms (Table 5). Scaling these technologies requires deploying and reading out large arrays of individual crystals or CCD chips, which presents significant technical challenges. However, their relatively small target masses are offset by their ability to achieve exceptionally low energy thresholds, enabling sensitivity to low-energy interactions that may remain inaccessible to larger-mass noble-liquid detectors [159].

Table 5. Comparison of detector target masses, fiducial or analysis masses, and representative exposures illustrating the scale and sensitivity of major direct DM detection experiments. Where a fixed fiducial mass is not defined, the analysis mass is reported as "analysis-dependent" because the effective target volume varies with event-selection criteria, energy range, and the specific physics analysis.

| Experiment | Active/ Target Mass [kg] | Fiducial/Analysis Mass [kg] | Latest Published/ Representative Exposure | Status | Ref. |
|---|---|---|---|---|---|
| ANAIS | 112.5 | 112.5 | 625.75 kg·yr | Data taking completed in Jan. 2026; six-year published analysis excludes the DAMA/LIBRA modulation at $4\sigma$; full-exposure analysis has been presented preliminarily | [32], [85] |
| ArDM | 850 | -- | ≈ 100 live days | Completed; ton-scale LAr technology demonstrator. Subsequent development transitioned toward the Global Argon Dark Matter Collaboration program | [160], [161] |
| CDEX | 10 | C10B-Ge1 | 205.4 kg·day | Active CDEX program at CJPL; ongoing low-threshold DM analyses | [89], [162], [163] |
| CoGeNT | 0.44 | ≈ 0.33 | 1129 live days (≈ 373 g·day) | Completed; data taking ended in 2013. Final 3.4-year analysis found no statistically significant evidence for a DM signal ($\sim 1.7\sigma$) | [35], [75] |
| COSINE | 106 | ≈ 61.3 | 358 kg.yr | Data taking at Y2L completed in March 2023; full 6.4-year dataset found no annual modulation and disfavors DAMA/LIBRA at $> 3\sigma$. Detector relocated to Yemilab and upgraded as COSINE-100U | [164] |
| COSINUS | ≈ 0.3 for the initial 8-module array | ≈ 0.3 | Target: 100 kg·day (Run 1) | Commissioning; Run 1 with eight NaI modules scheduled for 2026 | [165]–[167] |
| CRESST | 0.0236 | 0.0236 | 5.594 kg·day | Active; next-generation upgrade/data taking | [22], [37] |
| CYGNO | ≈ 0.6 | -- | No published DM-search exposure yet | CYGNO-04 (400 L) demonstrator under construction/ commissioning at LNGS | [38], [168] |
| DAMA/ LIBRA | ≈ 250 | ≈ 250 | 1.53 t·yr (phase2); 2.86 t·yr combined DAMA exposure | DAMA/LIBRA-phase2 completed; upgraded phase2-empowered configuration collected data with a lower threshold until fall 2024 | [84] |
| DAMIC | ≈ 0.04 | ≈ 0.04 | 11 kg·day | Original DAMIC program at SNOLAB transitioned toward skipper-CCD operation; successor DAMIC-M is under development/ deployment at LSM | [40], [113], [169], [170] |
| DarkSide | 46.4 | 31.3 | 16,660 kg.day | Completed; successor DarkSide-20k under development/ construction | [114], [171], [172] |
| DEAP | ≈ 3279 | ≈ 1266 | ≈ 2.74 t·yr | Active; detector upgrades aimed at reducing residual backgrounds and improving WIMP sensitivity | [58], [94], [116] |
| DMTPC | ≈ 0.0033 | ≈ 0.0033 | 35.7 g·day | R&D/demonstrator; directional $CF_4$ TPC technology demonstrated, with subsequent work focused on directional performance and scalability | [118], [173] |
| DRIFT | ≈ 0.14 (gas) | 0.0332 (F target) | ≈ 1.82 kg·day | Completed/R&D; background-free directional operation demonstrated | [95], [174] |
| EDELWEISS | ≈ 6.4 | -- (analysis-dependent) | 496 kg·day | EDELWEISS-III completed; ongoing EDELWEISS-SubGeV low-threshold R&D | [96], [175] |
| KIMS | 103.4 | 103.4 | 4,524.3 kg·day | Completed; CsI(Tl) phase concluded | [44], [176] |
| LUX (LZ) | 250 (7,000) | -- (analysis-dependent) | 10,091 kg·day (4.2 t.yr) | Completed; (active; accumulating data toward 1,000 live days) | [81], [177]–[179] |
| MIMAC | 6 L active volume | -- (analysis-dependent) | 495.1 and 354.2 live days (two chambers) | Active directional R&D; first directional DM limits reported | [45], [121] |
| NEWS-G | ≈ 0.114 (gas) | -- (analysis-dependent) | ≈ 1.1 kg.day | Active; S140/SNOGLOBE operated at LSM and subsequently deployed at SNOLAB | [146], [180] |
| NEWAGE | ≈ 0.10 (gas) | ≈ 0.10 | ≈ 3.18 kg.day | Active; underground directional DM search with 3D vector/head–tail reconstruction | [46], [123] |
| PandaX | ≈ 3,700 | 2,380 (Run0); 2,470 (Run1) | ≈ 1.54 t·yr | Operation completed Apr. 2026; transition to PandaX-20T | [100], [181] |
| PICO | 52 | ≈ 48 | 1404 kg·day | Completed; successor PICO-40L/PICO-500 program under development | [48], [182], [183] |
| PICOLON | 1.34 | 1.34 | 209 kg.day | Active R&D/scale-up; first annual-modulation search published; ultimate target ≈ 250 kg | [54], [184] |
| SABRE | 30 – 50 (planned) | -- | No DM search exposure yet | Construction/ commissioning at SUPL | [185], [186] |
| SENSEI | ≈ 0.1 | -- (analysis-dependent) | ≈ 0.10072 kg.day | Active at SNOLAB; upgraded Skipper-CCD array; sub-GeV DM searches ongoing | [88], [150], [187] |
| SuperCDMS | ≈ 30 (design payload) | -- (analysis-dependent) | No published SNOLAB DM-search exposure yet | First science data taking began Aug. 2026; initial science run underway | [72], [188] |
| TREX-DM | 0.30 (Ar) / 0.16 (Ne) | -- (analysis-dependent) | No published DM-search exposure yet | Active at LSC; high-pressure data taking and GEM + Micromegas development | [151], [152], [189] |
| XENON | ≈ 5,900 | -- (analysis-dependent) | 3.1 t·yr | Active; multi-tonne LXe TPC at LNGS | [102], [190], [191] |
| XMASS | 832 | 97 | 423 kg.yr | Completed; full dataset analyzed | [192], [193] |

*5.2. Energy Thresholds and Low-Mass Sensitivity*

The analysis threshold sets the minimum detectable DM mass. Because the recoil energy spectrum decreases exponentially, a lower threshold yields an exponential increase in signal acceptance for light WIMPs [194].

The precision frontier: Cryogenic bolometers and CCD-based detectors achieve energy thresholds in the eV to tens-of-eV range (Table 6). For nuclear-recoil searches, energy thresholds are commonly expressed in keV nuclear-recoil equivalent ($keV_{nr}$), which denotes the reconstructed energy of nuclear recoil and accounts for the reduced ionization or scintillation yield of nuclear recoils relative to electron recoils. By detecting phonons or single-ionization electrons, these instruments probe the sub-GeV DM mass regime. For example, Skipper CCDs can resolve individual electron events (~ 1.1 eV), enabling sensitivity to DM–electron scattering, an interaction channel inaccessible to conventional NR searches. Table 6 also shows that CRESST, NEWS-G, SuperCDMS, DAMIC, SENSEI, and TREX-DM are the principal experiments for the sub-10 GeV/$c^2$. The fourth column shows that SENSEI, DAMIC, CRESST, NEWS-G, TREX-DM, and SuperCDMS experiments currently provide the strongest reach into the sub-GeV DM mass range using different detection channels. While XENON, LUX, and PandaX probe down to a few GeV/$c^2$ through optimized S1/S2 analyses, they are primarily optimized for higher-mass WIMPs. Finally, COSINE, COSINUS, PICOLON, SABRE, and ANAIS are designed primarily to test the DAMA/LIBRA annual modulation signal rather than to achieve the lowest accessible DM masses.

The High-Energy Regime: XENON, LUX, and PandaX achieve the strongest SI exclusion limits near 30-50 GeV/$c^2$. Noble liquids and scintillating crystals typically operate at keV-scale thresholds. While sufficient for heavy WIMPs (>10 GeV), their efficiency decreases at lower masses because of the physics of signal generation (e.g., the energy required to produce a single scintillation photon) [63], [190].

Table 6. Representative energy thresholds of direct dark matter detector technologies and their implications for accessible dark matter mass ranges. Cells containing "--" indicate that no value is explicitly reported or that they are not applicable in the cited physics/performance analysis.

| Experiment | Numerical Threshold | Best Sensitivity Region / SI Mass Range [GeV/$c^2$] | Min Tested Mass [GeV/$c^2$] | Ref. |
|---|---|---|---|---|
| ANAIS | 1 $keV_{ee}$ | ≈ 8 and ≈ 54 (DAMA SI regions) | -- | [32], [195] |
| ArDM | 30 $keV_{nr}$ (Design goal) | -- | -- | [160], [161] |
| CDEX | 0.160 $keV_{ee}$ | ≈ 2 – 3 (light-mediator SI) | ≈ 2 | [162], [163] |
| CoGeNT | ≈ 0.4 $keV_{ee}$ | ≈ 7 – 11 | ≈ 5 | [75] |
| COSINE | ≈ 0.7 $keV_{ee}$ | ≈ 1 – 10 | ≈ 0.1 | [196], [197] |
| COSINUS | 1 keV (Projected NR threshold) | ≈ 1 – 10 (Projected SI sensitivity) | -- | [165], [167], [198] |
| CRESST | ≈ 6.7 eV | < 0.1 | < 0.1 | [22], [199] |
| CYGNO | ≈ 1 $keV_{nr}$ (Projected) | ≈ 1 – 10 (Projected) | -- | [38], [168], [200] |
| DAMA/LIBRA | 1 $keV_{ee}$ | ≈8 and ≈53 (canonical SI best-fit masses; model-dependent) | -- | [84], [86], [112], [201] |
| DAMIC | ≈ 50 $eV_{ee}$ | ≈ 1 – 10 (SI nuclear recoil) | ≈ 1 | [170], [202], [203] |
| DarkSide | ≈ 0.06 $keV_{er}$ | 1.2 – 3.6 | ≈ 1.2 | [172], [204] |
| DEAP | ≈ 50 $keV_{nr}$ | >30; (Optimum ≈ 100) | ≈ 30 | [94], [116] |
| DMTPC | ≈ 80 $keV_{nr}$ | ≈ 115 (SD proton) | -- | [173], [205] |
| DRIFT | ≈ 20 $keV_{nr}$ (Analysis dependent) | ≈ 100 (SD proton) | -- | [174] |
| EDELWEISS | ≈ 60 eV (Recoil energy) | 0.6 – 10 (SI; NR) | ≈ 0.6 | [96], [206], [207] |
| KIMS | 2 $keV_{ee}$ | ≤ 20 (SI); Low mass | -- | [97], [176] |
| LUX (LZ) | ≈ 1.1 $keV_{nr}$ (≈ 3 $keV_{nr}$) | ≈ 33 – 50 (≈ 30 – 50) | ≈ 4 (≈ 9) | [81], [98], [208], [209] |
| MIMAC | ≈ 2 $keV_{ee}$ | ≈ 10 (SD proton) | -- | [45], [121] |
| NEWS-G | 150 $eV_{ee}$ | ≈ 0.17 – 1.2 (SD proton) | ≈ 0.17 | [146], [180] |
| NEWAGE | ≈ 50 $keV_{ee}$ | ≈ 150 (SD proton) | ≈ 10 | [46], [123] |
| PandaX | ≈ 0.33 $keV_{nr}$ | 2.5 – 5.0 (Low-mass SI) | ≈ 2.5 | [100], [210] |
| PICO | 2.45 keV (Thermodynamic) | ≈ 25 (SD proton) | ≈ 3 (for special low-threshold runs) | [182], [211] |
| PICOLON | ≈ 2 – 6 $keV_{ee}$ (Analysis ROI) | ≈ 100 (Reported SI benchmark) | -- | [54], [148] |
| SABRE | ≈ 1 – 2 $keV_{ee}$ (Design) | ≈ 30 (Projected SI) | -- | [149], [186], [212] |
| SENSEI | ≈ 3.75 eV | ≈ 0.0005 – 1 (DM-electron) | ≈ 0.0005 | [50], [88], [150] |
| SuperCDMS | ≈ 40 $eV_{nr}$ (Ge HV; Projected) | ≈ 1 – 5 (SI, NR, projected) | ≈ 0.3 (Projected) | [62], [101], [213], [214] |
| TREX-DM | ≈ 1 $keV_{ee}$ (Achieved) | ≈ 2 – 8 (Projected SI) | -- | [151], [215] |
| XENON | ≈ 3.8 $keV_{nr}$ | ≈ 30 – 41 | ≈ 10 | [190], [216], [217] |
| XMASS | ≈ 1.0 $keV_{ee}$ | ≈ 60 | ≈ 4 | [192], [218] |

### *5.3. Backgrounds*

The sensitivity of direct DM experiments depends not only on achieving low energy thresholds but also on minimizing backgrounds and distinguishing signal-like events from radioactive instrumental backgrounds, as listed in Table 7. ERs, produced mainly by $\beta$ and $\gamma$ radiation, constitute an important background in many detectors, while neutron-induced nuclear recoils are particularly challenging because they can closely mimic DM-induced NRs. Different detector technologies therefore employ distinct background-discrimination strategies:

- Cryogenic semiconductor detectors, such as SuperCDMS and EDELWEISS, measure phonon signals together with ionization or scintillation channels, allowing ER/NR discrimination while maintaining very low recoil-energy thresholds. Related cryogenic calorimeters such as CRESST combine phonon and scintillation measurements to distinguish nuclear recoils from electromagnetic backgrounds.
- Pure ionization detectors, including low-threshold Ge experiments such as CoGeNT and CDEX, rely primarily on charge collection and therefore have more limited event-by-event ER/NR discrimination. Background suppression relies on pulse-shape analysis, surface-event rejection, shielding, radiopure materials, and detailed spectral modeling.
- Bubble chambers, such as PICO, provide strong intrinsic rejection of ER backgrounds because bubble nucleation is highly suppressed for low-density electronic energy deposition, while acoustic information can help distinguish nuclear recoils from $\alpha$-induced events.
- Noble-liquid TPCs, including XENONnT, LZ, PandaX, and DarkSide, combine scintillation and ionization signals with three-dimensional event reconstruction, fiducialization, and self-shielding to suppress ER, surface, and external backgrounds. Liquid-argon detectors additionally exploit pulse-shape discrimination, while single-phase experiments such as DEAP rely primarily on scintillation timing and fiducial-volume selection.
- Scintillating-crystal detectors, such as DAMA/LIBRA, ANAIS, COSINE, PICOLON, SABRE, and KIMS, generally have more limited event-by-event ER/NR discrimination and therefore rely strongly on radiopure materials, shielding, environmental monitoring, coincidence rejection, and, in modulation searches, the characteristic time dependence of a possible DM signal.
- CCD and imaging detectors, including DAMIC and SENSEI, achieve exceptionally low ionization thresholds and use spatial charge topology, detector depth information, shielding, and material radiopurity to suppress backgrounds, although their event-by-event ER/NR discrimination is more limited than in dual-channel detectors.

Table 7. Representative background levels in selected low-energy signal regions for major direct DM experiments, together with the corresponding energy ranges used for background evaluation and detector-specific background rejection or discrimination methods.

| Experiment | Background Level [events/(kg·day·keV)] | Energy Range | Background Rejection / Discrimination Method | Ref. |
|---|---|---|---|---|
| ANAIS | ≈ 3.58 | 1 – 6 $keV_{ee}$ | Pulse-shape/noise filtering, anticoincidence between NaI(Tl) modules, muon veto, shielding, and background modeling | [139], [219] |
| ArDM | --[a] | Low-energy region, <1 MeV; WIMP sensitivity at tens of $keV_{ee}$ | Pulse-shape discrimination, event localization/fiducialization, ionization-density information, and shielding | [160] |
| CDEX | ≈ 2.5 | 2 – 4 $keV_{ee}$ | Noise/pulse-shape cuts, bulk–surface event discrimination, fiducialization, and passive shielding | [163] |
| CoGeNT | ≈ 3 | 0.5 – 3.0 $keV_{ee}$ | Pulse-rise-time discrimination for surface events, shielding, and spectral background modeling | [75] |
| COSINE | $3.167 \pm 0.022$ (Observed) | 0.7 – 6 $keV_{ee}$ | Pulse-shape/noise cuts, multiple-hit rejection, liquid-scintillator veto, shielding, and background modeling | [220]–[222] |
| COSINUS | $\sim 1$ (Design assumption) | ≥ 1 keV; 2 – 6 $keV_{ee}$ for DAMA comparison | Event-by-event discrimination using Phonon/light ratio | [36], [223] |
| CRESST | ≈ 2.2 | 1 – 140 keV | Phonon/light ratio, scintillating housing, instrumented holders, and event-quality cuts | [37], [224], [225] |
| CYGNO | Design-dependent / projected | Few keV– tens of $keV_{ee}$; NR threshold targeted around ~ 1 keV | 3D track topology, ionization-density information, and optical imaging; directional reconstruction provides additional background rejection | [38], [141], [168] |
| DAMA/LIBRA | ≈ 1 | 2 – 6 $keV_{ee}$ | Radiopure NaI(Tl) crystals, passive shielding, multiple-hit rejection, noise/pulse-shape cuts, and annual-modulation analysis | [226] |
| DAMIC | ≈ 3.1 (counts/kg.day.$keV_{ee}$) | 2.5 – 7.5 $keV_{ee}$ | Event topology, depth reconstruction, surface-event rejection, and energy information | [40], [227], [228] |
| DarkSide | ≈ 1.5 | 0.5 $keV_{ee}$ (representative point; analysis fully efficient above 0.1 $keV_{ee}$) | S1/S2 information, pulse-shape discrimination, fiducialization, single-scatter selection, and active neutron/muon vetoes | [204] |
| DEAP | $\approx 4.7 \times 10^{-8}$ | 15.6 – 32.9 $keV_{ee}$ | Pulse-shape discrimination, fiducialization, position reconstruction, and dedicated α/Cherenkov rejection cuts | [229]–[231] |
| DMTPC | ≈ 24.5 | 80 – 200 keV | Recoil-track topology and direction, CCD–charge coincidence, and veto of α/surface events | [173], [205] |
| DRIFT | 0 Observed after cuts | 700 – 6000 NIPs (recoil-energy conversion is nucleus dependent) | 3D fiducialization, recoil-track topology/direction, ionization-density discrimination, minority-carrier timing, and neutron shielding | [174] |
| EDELWEISS | ≈ 0.08 – 0.24 | 20 – 200 keV | Ionization yield + interleaved veto electrodes; fiducial-volume selection | [175], [232]–[234] |
| KIMS | ≈ 2 – 3 | 2 – 4 $keV_{ee}$ | Pulse-shape discrimination, single-hit selection, surface-α rejection, shielding, and internal-background reduction | [44], [176] |
| LUX | $\approx (6.3 \pm 0.5) \times 10^{-5}$ (Observed) | < 15 $keV_{ee}$ | S2/S1 discrimination, 3D fiducialization, single-scatter selection, self-shielding, and active xenon-skin + outer-detector vetoes | [120], [235], [236] |
| MIMAC | --[b] | ≈ 4.5 – 10 $keV_{ee}$ | 3D track topology + BDT electron-recoil rejection + recoil direction | [121] |
| NEWS-G | ≈ 1.8 (Projected) | 0.014 – 1 $keV_{ee}$ | Pulse-shape / rise-time discrimination + fiducialization | [180] |
| NEWAGE | ≈ 0.23 (Expected) | 50 – 60 $keV_{ee}$ | Track length and topology, fiducialization, and recoil-direction reconstruction | [46], [123] |
| PandaX | $\approx (4.9 \pm 0.5) \times 10^{-5}$ | 1- 10 $keV_{ee}$ | S2/S1 discrimination, 3D fiducialization, single-scatter selection, self-shielding, and veto cuts | [210], [237], [238] |
| PICO | $\approx$ (1 $\pm$ 0.4 over 1404 kg.day exposure) | 2.45 $keV_{nr}$ | Thermodynamic ER rejection + acoustic parameter + bubble multiplicity + fiducial/optical cuts | [182], [239], [240] |
| PICOLON | ≈ 1.5 (Observed) | 2 – 6 $keV_{ee}$ | Pulse-shape/noise cuts + radiopure NaI(Tl) + anticoincidence | [54] |
| SABRE | 0.72 (Projected) | 1 – 6 $keV_{ee}$ | Radiopure NaI(Tl) + active liquid-scintillator veto + passive shielding + muon veto | [149], [212] |
| SENSEI | --[c] | Discrete 2 – 10 $e^-$ charge bins | Pixel topology + Skipper-CCD single-electron charge resolution + spatial/temporal masking | [50], [88] |
| SuperCDMS | $\approx 1.3 \times 10^{-3}$ (Expected post-selection) | 1.6 – 10 $keV_{nr}$ | Ionization/phonon ratio + phonon timing + fiducialization | [72], [101], [241] |
| TREX-DM | ≈ 80 (Observed) | 0.2 – 7 $keV_{ee}$ | Cluster topology + event morphology + fiducial cuts + radiopure Micromegas | [55], [242], [243] |
| XENON | $\approx (4.3 \pm 0.4) \times 10^{-5}$ | 1 – 30 $keV_{ee}$ | S2/S1 ratio + 3D fiducialization + single-scatter selection + active vetoes | [190], [244], [245] |
| XMASS | $\approx (4.2 \pm 0.2) \times 10^{-3}$ (Observed) | 5 $keV_{ee}$; WIMP analysis: 2 – 15 $keV_{ee}$ | Fiducialization + scintillation timing/pattern cuts + surface-event rejection + self-shielding | [192], [218] |

[a] ArDM: no single directly comparable spectral background rate is reported; the dominant intrinsic ER background corresponds to a measured $^{37}Ar$ activity of $0.95 \pm 0.05$ Bq/kg.

[b] MIMAC: background is evaluated data-driven using directional ON/OFF regions; a directly comparable differential background rate is not reported.

[c] SENSEI: backgrounds are reported in discrete electron-multiplicity bins rather than as a conventional differential background rate.

- Gaseous detectors suppress backgrounds through ionization topology, drift information, pulse-shape analysis, fiducialization, and radiopure detector construction. Spherical proportional counters such as NEWS-G emphasize very low thresholds and surface-event rejection, while *directional gaseous detectors*, including DRIFT, DMTPC, MIMAC, NEWAGE, and CYGNO, additionally reconstruct recoil-track morphology and direction to distinguish nuclear recoils from backgrounds and search for the expected anisotropic signature of Galactic DM.

*5.4. Current Exclusion Limits*

The global landscape of exclusion limits falls into two primary interaction types: spin-independent (SI) and spin-dependent (SD).

Derivation of Exclusion Limits: In direct-detection experiments, exclusion limits are obtained by comparing the observed recoil-energy distribution with the expected signal and background models. For a given DM mass and interaction cross section, the expected event rate is calculated from the assumed DM velocity distribution, local DM density, target composition, and detector response. Statistical methods, such as profile-likelihood analyses, Bayesian inference, or Poisson-based upper limits, then determine the largest signal contribution compatible with the observed data at a specified confidence level, commonly 90%. Repeating this procedure over a range of DM masses produces an exclusion curve in the DM mass-cross-section parameter space [20], [246].

SI interactions: The SI interactions benefit from coherent scattering across the nucleus, with the nuclear cross section scaling approximately as $A^2$, where $A$ is the nuclear mass number. This coherent enhancement favors heavy nuclei. As shown in Figure 3, liquid-xenon experiments such as XENONnT, and PandaX-4T provide stringent constraints across much of the intermediate- and high-mass WIMP region, whereas cryogenic and semiconductor detectors become increasingly important at lower masses because of their lower recoil-energy thresholds.

- High-mass DM ($\gtrsim$ 10 GeV): Liquid-xenon TPCs, such as XENONnT, LZ, and PandaX-4T, benefit from the coherent enhancement of xenon nuclei combined with large target masses and multi-tonne-scale exposures.
- Low-mass DM ($\lesssim$ 10 GeV): Cryogenic detectors, such as SuperCDMS, and CCD-based experiments, such as DAMIC, achieve sensitivity to lower WIMP masses through their ultra-low energy thresholds. Although Ge and Si targets provide weaker coherent enhancement than xenon, their low thresholds enable detection of the small recoil energies expected from light WIMPs [247].

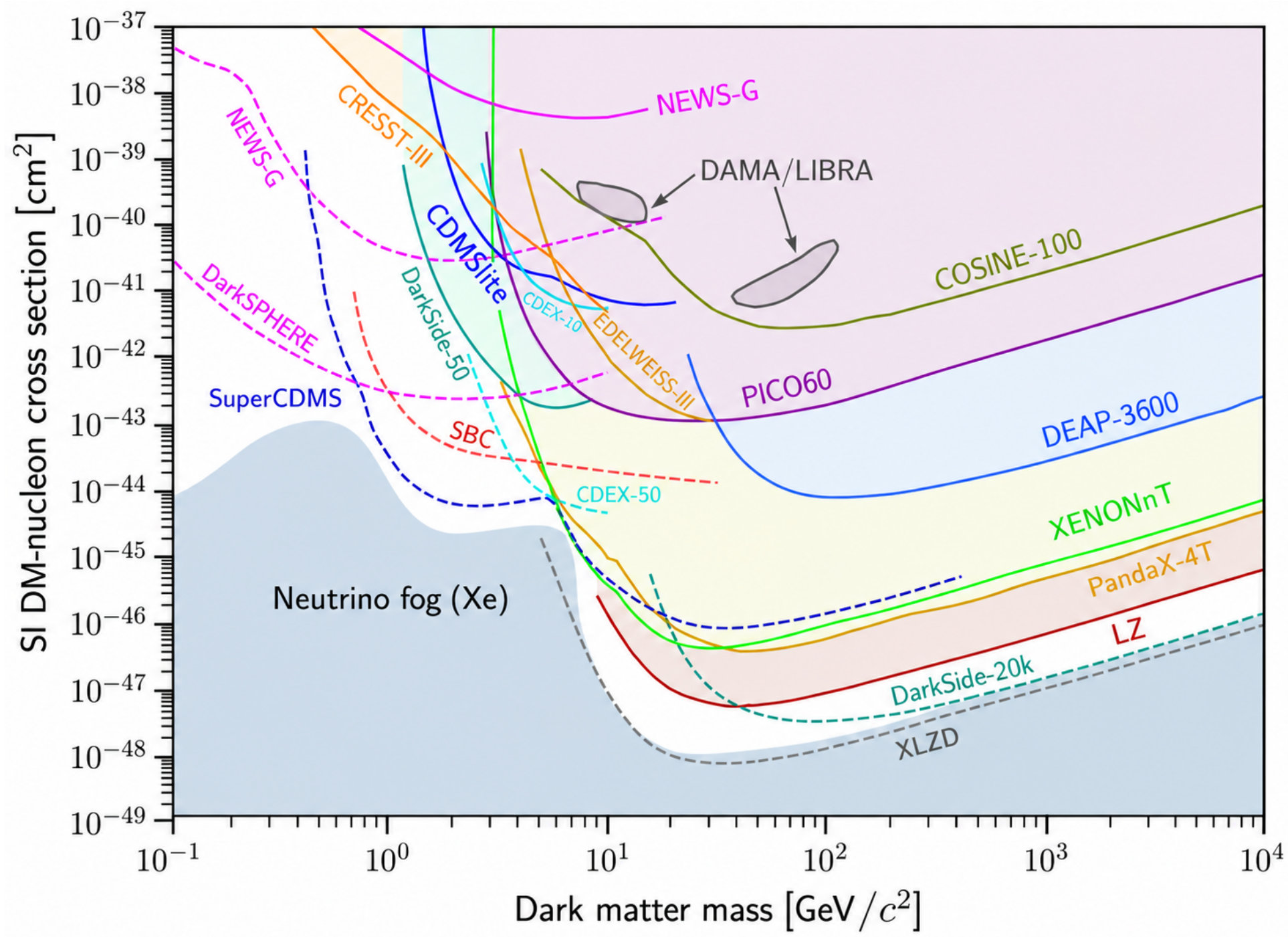

Figure 3. Experimental spin-independent dark matter-nucleon scattering cross-section limits as a function of dark matter mass. Solid curves show current direct-detection limits at 90% C.L., while dashed curves indicate projected sensitivities at 90% C.L. Adapted from Ref. [158].

SD interactions: In contrast to SI scattering, SD interactions couple primarily to nuclear spin ($J$), arising from the net spin contribution of the unpaired nucleons rather than coherent coupling to the entire nucleus [248], [249]. Sensitivity therefore depends strongly on the target's spin structure and isotopic composition. The left panel of Figure 4 shows the SD neutron-coupling limits, where liquid-xenon experiments such as LZ, and XENONnT are particularly competitive because naturally occurring xenon contains the odd-neutron isotopes $^{129}Xe$ and $^{131}Xe$, both of which possess non-zero nuclear spin.

The right panel of Figure 4 presents the SD proton-coupling limits, for which bubble-chamber experiments, particularly PICO, provide the strongest sensitivity. Fluorocarbon targets such as $C_3F_8$ contain abundant $^{19}F$, whose nuclear spin is dominated by an unpaired proton, making fluorine-based detectors especially effective for SD-proton searches.

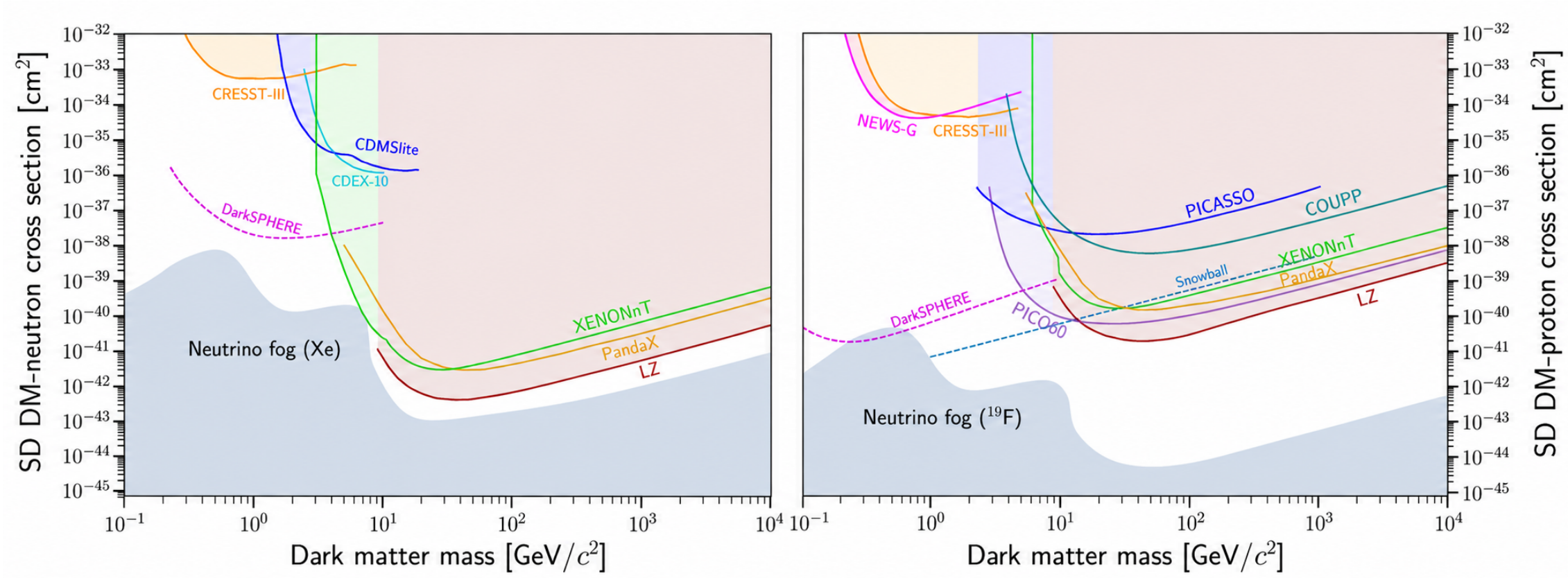

Figure 4: Experimental spin-dependent dark matter–nucleon scattering cross-section limits as a function of dark matter mass at 90% C.L. *Left panel*: SD dark matter–neutron scattering limits. *Right panel*: SD dark matter–proton scattering limits. Adapted from Ref. [158].

### *5.5. Specialized Observables: Directionality and Modulation*

While most experiments count total events, a subset focuses on unique kinematic signatures to provide "smoking-gun" evidence of DM.

Directionality: Directional detection provides a distinctive way to establish the Galactic origin of a DM signal by exploiting the anisotropic nuclear-recoil distribution expected from the apparent DM wind associated with the Solar System's motion through the Galactic halo [70], [250]. Low-pressure gaseous TPCs, such as DRIFT and CYGNUS, reconstruct the topology and direction of nuclear recoils, allowing the expected recoil anisotropy associated with Cygnus direction to be distinguished from approximately isotropic terrestrial [70], [250], [251]. Directional sensitivity is, however, not restricted to gaseous detectors. Nuclear-emulsion detectors, particularly the NEWSdm project at the Gran Sasso National Laboratory (LNGS), employ nanometric-grain emulsions as both target and tracking medium to record sub-micrometer nuclear-recoil tracks. NEWSdm uses Nano Imaging Tracker emulsions and an equatorial mounting concept to preserve the detector orientation relative to the expected Galactic DM-wind direction, providing a solid-state alternative to gaseous directional TPCs [252]. Paleodetectors offer a complementary concept, in which nanometer-scale damage tracks produced by nuclear recoils are retained in ancient minerals over geological timescales. Their enormous effective exposure and track-based readout could extend sensitivity to DM interactions beyond conventional real-time detectors, although reconstructing and controlling radiogenic, cosmogenic, and geological backgrounds remain a central challenge [253].

Annual modulation: Scintillating-crystal experiments, most notably DAMA/LIBRA, search for the expected few-percent annual modulation in the event rate arising from the annual variation of the Earth's velocity relative to the Galactic DM halo [83], [254], [255]. DAMA/LIBRA reports a persistent and statistically significant modulation in its NaI(Tl) detectors; however, ANAIS-112 and COSINE-100, which employ the same NaI(Tl) target material, have tested this interpretation independently and provide comparatively model-independent tests of the DAMA/LIBRA result. The six-year ANAIS-112 analysis found no significant annual modulation and reported incompatibility with the DAMA/LIBRA signal at approximately $4\,\sigma$ [85]. Similarly, the full COSINE-100 dataset shows no evidence for a compatible annual modulation and disfavors the DAMA/LIBRA interpretation at greater than $3\,\sigma$, depending on the energy calibration and recoil-energy representation [164]. A recent combined analysis of ANAIS-112 and COSINE-100 further strengthens this tension: using their six-year datasets, the combined modulation amplitudes are consistent with zero and exclude the DAMA/LIBRA modulation at $4.7\sigma$ in 1 – 6 keV and $3.5\,\sigma$ in 2 – 6 keV [256]. These same-target measurements therefore place strong experimental constraints on a DM interpretation of the DAMA/LIBRA modulation, although differences in detector response, calibration, and systematic effects remain relevant when comparing the experiments.

### *5.6. Summary of the Experimental Landscape*

The historical trajectory of direct detection reflects a progression from small-scale prototypes toward large, coordinated observatories as summarized in Table **8**. The field has now entered a stage of strategic complementarity, in which distinct detector technologies occupy separate regions of parameter space rather than competing directly. Noble-liquid TPCs provide broadband sensitivity and remain the leading technology for the high-mass WIMP searches, owing to their scalability to multi-tonne target masses. Cryogenic bolometers and skipper-CCD detectors serve as precision instruments for the low-mass and sub-GeV regime, where ultra-low energy thresholds are essential. Bubble chambers and gaseous projection chambers occupy a complementary niche, offering sensitivity to SD coupling and, in directional detectors, a potential means of confirming a galactic origin for any candidate signal. A comparative sensitivity of some direct DM detection experiments is depicted in Figure 5. This technological diversity is not incidental but a deliberate feature of the field's maturation: should a signal emerge, it can be cross-checked across multiple target nuclei, interaction channels, and systematic-error profiles, providing a robust test of its particle-physics origin.

Table 8. Operational timelines of major dark matter experiments.

| Experiment | Start of Major Runs | Ref. |
|---|---|---|
| ANAIS | 2017 | [139] |
| CDEX | ≈ 2010 | [34] |
| CoGeNT | ≈ 2008 | [75] |
| COSINE | ≈ 2016 | [256] |
| COSINUS | 2024 (commissioning) | [36] |
| CRESST | 2016 | [22] |
| CYGNO | 2022 (prototype) | [141] |
| DAMA/LIBRA | ≈ 1996 | [201] |
| DAMIC | ≈ 2012 | [40], [92] |
| DEAP | 2016 – 2017 | [58] |
| DMTPC | ≈ 2010 | [118] |
| DRIFT | ≈ 2001 | [42] |
| EDELWEISS | ≈ 2014 | [96] |
| KIMS | ≈ 2000 | [44] |
| LUX (LZ) | 2013 (2022) | [52], [235] |
| MIMAC | ≈ 2010 | [257] |
| NEWS-G | 2017 | [47] |
| NEWAGE | ≈ 2008 | [46] |
| PandaX | II - 2014<br>4T - 2021 | [258] |
| PICO | ≈ 2013 | [48] |
| PICOLON | 2023 (pilot) | [54] |
| SABRE | 2023 (commissioning) | [49] |
| SENSEI | 2018 | [50] |
| SuperCDMS | Soudan - 2012;<br>SNOLAB - 2026 | [62], [72] |
| TREX-DM | ≈ 2018 | [55] |
| XENON | 1T - 2016<br>nT - 2021 | [129], [157] |

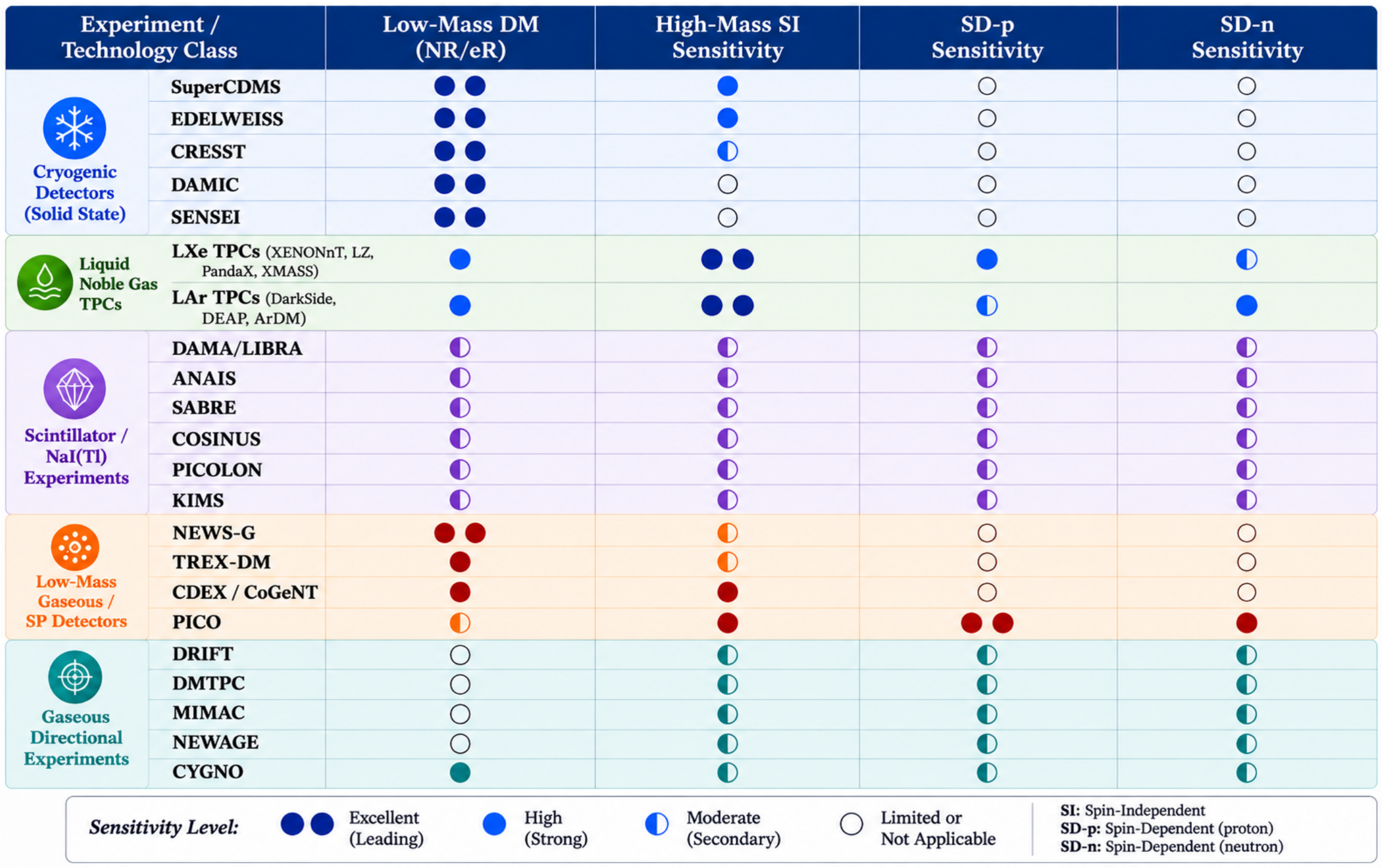

Figure 5: A condensed matrix to highlight the three principal performance metrics: low-mass sensitivity, high-mass SI sensitivity, SD proton and SD neutron sensitivity. Image resources: [7], [8], [63], [249], [251], [259], [260]

## 6. Future Directions and Outlook

The next phase of direct DM detection is defined by the approach toward the neutrino-limited regime. The irreducible background arises from coherent elastic neutrino–nucleus scattering ( CE$\nu$NS ). As we approach this limit, conventional background-reduction strategies become ineffective. Future experiments, therefore, pursue three complementary strategic objectives: scaling target masses to maximize exposure before reaching this boundary, lowering energy thresholds to probe light DM candidates, and developing novel observables that can statistically or experimentally distinguish WIMP-induced signals from neutrino interactions.

### 6.1. *Scaling to the Neutrino Boundary*

For high-mass WIMP searches, the primary goal is to maximize exposure while maintaining background-free operation. Next-generation noble-liquid experiments (Table 9) are targeting the multi-tonne scale and beyond.

- Liquid Argon: Experiments such as DarkSide-20k aim to exploit argon's excellent PSD to suppress ER backgrounds to the level required to probe WIMP-nucleon cross sections approaching the neutrino floor [82]. The Global Argon Dark Matter Collaboration (GADMC) is pursuing a longer-term program in which ARGO, envisioned as a large-scale successor to DarkSide-20k, will further extend the sensitivity of liquid-argon detectors towards and beyond the neutrino floor. ARGO is currently being developed as a concept with an approximately 300-tonne fiducial liquid-argon target, with an exposure goal of roughly 3000 tonne-years [261].

- Liquid xenon: Multi-tonne detectors (e.g., PandaX-xT, Darwin) rely on self-shielding and extreme radiopurity. As these detectors grow, solar and atmospheric neutrinos will eventually become the dominant background. At this stage, simple "counting" experiments lose sensitivity, as the neutrino rate becomes indistinguishable from a WIMP signal based solely on the energy spectrum [262].

Table 9. An overview of selected next-generation and emerging direct DM detection experiments and concepts, highlighting their projected scale and primary scientific objectives.

| Experiment / Concept | Planned scale/ Development scale | Aim | Ref. |
|---|---|---|---|
| ANAIS+ | NaI(Tl)+SiPM prototype at ~100 K | Improved sensitivity to low-mass/SD-proton DM and a more robust DAMA/LIBRA test | [139] |
| ASTAROTH | ≈ 360 g NaI(TI) prototype coupled to a $5 \times 5\ cm^2$ SiPM matrix at ≈ 80 K | Demonstrate cryogenic SiPM readout of NaI(Tl) for lower thresholds and reduced backgrounds in future DM searches | [263] |
| BULLKID | Prototype modules (scalable to ≈ 10 kg) | Kinetic-inductance cryogenic detectors for low-energy NR | [264], [265] |
| COSINE | 99.1 kg NaI(Tl), 8 upgraded crystals | Lower-threshold low-mass DM search and DAMA/LIBRA test | [266] |
| COSINUS | 8×34.8 g initially; 20× ~100 g planned | Model-independent test of DAMA/LIBRA using cryogenic NaI | [167] |
| CRESST | Modular, gram scale detectors; future scaling under development | Ultra-low threshold cryogenic search for sub-GeV DM | [199], [267] |
| CYGNO | Multi-m$^3$ gaseous TPC program | Directional and low-mass DM searches | [200] |
| DAREDEVIL | Gram-scale prototype | eV/meV-scale calorimetry for sub-GeV DM electron scattering | [268] |
| DarkSide | ≈ 20 tonnes LAr | High exposure WIMP search approaching | [269] |
| | | neutrino background regime | |
| DARWIN | ≈ 40 tonnes active LXe | Multi-tonne liquid-xenon observatory for WIMP DM, neutrino physics and rare-event searches | [270] |
| DARKSPHERE | 1.4 m-diameter spherical proportional counter | Low-mass DM using ultra-low-threshold gaseous detection | [271] |
| DeLIGHT | Prototype / gram-scale R&D | Superfluid - $^4He$ detection of sub-GeV DM | [272] |
| (CYGNUS-style networks) | Multi-$m^3$ directional gas detectors | Directional detection and neutrino floor studies | [273] |
| DEAP-50T / Future large LAr detector | $\mathcal{O}$(50 tonnes) | PSD-based multi-tonne program | [56] |
| NEWS-G | ≈ 140 kg (SNOLAB concept) | Low-mass DM with light noble-gas targets | [180] |
| NEWAGE | Multi-$m^3$ gaseous TPC concept | Directional DM detection | [124] |
| OSCURA | 10 kg Si skipper-CCD array; target exposure ≈ 30 kg.yr at SNOLAB | Ultra-low-threshold search for sub-GeV DM with single-/few-electron sensitivity | [274], [275] |
| PandaX-XT | ≈ 43 tonnes active LXe | High-statistics SI DM searches | [276]–[278] |
| PICO-500 | ≈ 250 L $C_3F_8$ bubble chamber; ≈ 250 kg target | Next-generation bubble-chamber search, particularly for SD proton-coupled DM | [279], [280] |
| PICOLON | ≈ 250 kg (planned upgrade) | Ultra-radiopure NaI(TI) detector for annual modulation | [54] |
| ResNOVA | ≈ 100 kg – 1 tonne concept | Cryogenics $PbWO_4$ detector targeting CE$\nu$NS and light DM | [281], [282] |
| SABRE | ≈ 50 kg NaI(TI) | Independent test of DAMA annual modulation | [186], [283] |
| SBC | Two 10 kg LAr chambers | Low threshold GeV-scale DM and CE$\nu$NS | [66] |
| SENSEI | ≈ 10 kg (future skipper CCD arrays) | Large-scale ER search from MeV – GeV DM | [88] |
| SuperCDMS SNOLAB | ≈ 30 kg (current), up to ≈ 100 kg (future concept) | Approach the neutrino floor for 1 – 10 GeV DM | [62], [188] |
| TESSERACT | ≈ 10 – 100 kg (modular) | Milti-technology search for sub-GeV DM using skipper CCDs, TESs, and cryogenic sensors | [284] |
| TREX-DM | ≈ 10 – 20 kg (future high-pressure TPC concept) | High-pressure gaseous TPC optimized for low-mass WIMPs | [215] |

### *6.2. The Low-Mass Frontier*

In parallel, the search for Sub-GeV DM drives the development of modular arrays rather than monolithic volumes. Here, the engineering challenge is not only mass but also readout stability and noise performance.

- Cryogenic Arrays: Projects such as SuperCDMS SNOLAB are scaling cryogenic Ge and Si detector arrays while using high-voltage Neganov–Trofimov–Luke (NTL) amplification to enhance phonon signals and push recoil thresholds toward the eV scale [285]. Complementarily, the TESSERACT program is developing sub-eV-resolution cryogenic detectors based on transition-edge sensors and multiple target materials, including polar and scintillating crystals in SPICE and superfluid helium in HeRALD, to extend direct searches to MeV–GeV and lighter DM candidates. Its multi-target strategy is also designed to help distinguish nuclear- and electron-recoil signals from instrumental low-energy backgrounds [284], [286] .
- Skipper CCDs: Next-generation silicon sensors (e.g., OSCURA) aim to deploy kilogram-scale arrays of Skipper CCDs that provide sub-electron-noise levels. This enables unambiguous identification of single-electron events, which is essential for probing DM-electron scattering models.

### *6.3. New Observables: Directionality*

To advance beyond the neutrino-limited regime, experiments must measure more than energy deposition. Directionality is the only known signature capable of distinguishing WIMPs (coming from the Cygnus constellation) from solar neutrinos (coming from the Sun) or atmospheric neutrinos (isotropic) [287].

While current directional detectors based on gas TPCs are limited in achievable target mass, future concepts, such as those proposed by the CYGNUS Collaboration, envision large-scale, modular gas detector networks with total volumes of $10$ -$100\,\mathrm{m}^3$. Although these detectors cannot match noble-liquid experiments in raw exposure, their unique ability to reconstruct NE directions makes them indispensable verification instruments, particularly if a signal is claimed in the neutrino-dominated regime.

### *6.4. The AI Revolution in Analysis*

As detector complexity increases, direct DM searches are increasingly moving beyond conventional cut-based selections toward integrated ML frameworks. These approaches are particularly attractive for low-threshold experiments, where distinguishing physical events from detector noise can depend on subtle correlations across multiple channels, pixels, or time samples.

Modern ML techniques, including Convolutional Neural Networks (CNNs), boosted decision trees, and emerging graph-based approaches, are being investigated for several aspects of direct-detection analysis:

1. Maximize Information Extraction: Rather than reducing detector data to a small set of summary quantities, ML algorithms can operate directly on raw or minimally processed waveforms, multidimensional detector images, or correlated readout channels. In cryogenic semiconductor detectors, for example, researchers have investigated CNNs to reconstruct event energies directly from phonon waveforms. SuperCDMS has subsequently developed a multi-channel, multi-template reconstruction framework using ML to account for correlations between readout channels and position-dependent pulse shapes, improving the extraction of energy and position information from complex phonon signals [107]. This shows how ML can exploit information that conventional pulse-processing methods may lose when they reduce the waveform to a small set of parameters.
2. Image-and Topology-Based Classification: Imaging detectors provide a natural application for ML because individual interactions are represented as spatially correlated charge patterns. CCD experiments such as DAMIC and SENSEI already exploit pixel-level information, cluster morphology, and spatial correlations to identify and reject instrumental backgrounds. These data structures can also be used as inputs to image-classification algorithms, including CNNs, although the published DM analyses generally retain physically interpretable event-selection and masking procedures rather than relying exclusively on deep-learning classifiers [65], [150].
3. Multivariant Background Discrimination: ML classifiers can combine multiple observables that individually provide only modest discrimination. In noble-liquid and cryogenic detectors, relevant features can include pulse shape, S1/S2 information, phonon-channel correlations, reconstructed position, and event-quality parameters. Boosted decision trees and related classifiers can therefore provide a more flexible representation of signal and background populations than one-dimensional cuts. Importantly, ML should complement rather than replace physically motivated observables and statistical inference.
4. Physics-Informed and Hybrid ML: A major challenge in applying ML to rare-event searches is the potential mismatch between training on simulations, which may learn imperfections in the detector model, and training on calibration data, which may not generalize to a low-background search region. Consequently, an increasingly important strategy is hybrid or physics-informed ML, in which machine-learning models are constrained or validated using detector or response models, calibration data, control samples, and physically motivated simulations. This approach helps mitigate dataset-shift effects and provides physically interpretable validation of the reconstructed observables.

5. Validation and Statistical Robustness: Because DM searches often operate in regimes with very small, expected signal rates, apparently modest ML biases can have significant consequences for exclusion limits or discovery claims. ML-based reconstruction and classification should therefore be validated using statistically independent calibration data, background control samples, and dedicated simulation studies. In particular, demonstrate the classifier's stability against changes in detector conditions, noise distributions, and calibration datasets before incorporating it into the final statistical inference. The recent SuperCDMS reconstruction study provides an instructive example: although ML can improve reconstruction in complex multi-channel data, its performance must be evaluated under variations in the underlying noise and detector response [107].

Overall, AI's role in direct DM detection is shifting from an auxiliary classification tool to an integrated component of event reconstruction, feature extraction, and statistical inference. Nevertheless, the extreme rarity of potential signals makes transparency, independent validation, and control of systematic uncertainties essential. The most promising future direction is therefore not a purely black-box analysis, but a hybrid framework in which ML extracts high-dimensional information while the final interpretation remains anchored to detector physics and well-defined statistical models.

## 7. Conclusions

Direct DM detection has evolved from small-scale prototypes into a precision discipline operating at the fundamental limits of radiometric quiet. The current experimental landscape is marked by strategic complementarity: Noble Liquid TPCs drive high-mass sensitivity, Cryogenic and CCD detectors explore the low-mass frontier, and specialized technologies target SD and directional signatures.

Despite dramatic sensitivity improvements spanning orders of magnitude over the last two decades, the particle nature of DM remains elusive. The absence of a discovery to date does not imply failure; rather, it has systematically narrowed the parameter space and motivated expansion into new regimes, particularly the sub-GeV domain.

A particularly important unresolved issue is the DAMA/LIBRA annual-modulation signal, which has motivated extensive experimental and theoretical efforts for more than two decades. The central question is whether the observed modulation is caused by DM or by an unaccounted-for time-dependent background. The DAMA/LIBRA annual modulation puzzle is being investigated by several independent NaI(TI)-based experiments. ANAIS and COSINE have accumulated more than six years of data and reported annual-modulation searches that find no evidence supporting the DAMA/LIBRA modulation signal. Complementary next-generation efforts, including SABRE and PICOLON, are pursuing improved radiopurity, detector stability, and environmental background control to provide further independent tests of the DAMA/LIBRA result. Emerging comparisons of results from these experiments, together with the development of next-generation NaI(Tl) detectors with improved sensitivity and background control, provide the most direct route to resolving the discrepancy. Thus, the DAMA/LIBRA puzzle is increasingly being transformed from a single-experiment anomaly into a testable question of reproducibility and systematic control.

The field's future will not be defined by a single "winner" technology. Instead, it will rely on a coordinated, multi-messenger, multi-target approach. The field is now entering the neutrino-limited regime, rather than merely approaching it: the 2024 XENONnT and PandaX-4T results provide the first indicators of solar $^{8}B$ neutrino-induced nuclear recoils through coherent elastic neutrino-nucleus scattering ($\mathrm{CE}\nu\mathrm{NS}$) in DM detectors. This milestone shifts the role of neutrinos from a projected sensitivity limit to an experimentally relevant background that must be measured, modeled, and incorporated into DM inference. Future experiments will therefore increasingly operate as both DM and neutrino observatories, exploiting larger exposures and improved event reconstruction to distinguish potential DM signals from neutrino-induced recoils. In parallel, AI-assisted analysis and complementary target materials may provide additional information to separate these populations and extend sensitivity beyond the conventional neutrino-limited regime.

## Acknowledgements

The College of Natural and Health Sciences at Zayed University partially supported this review paper through Dr. Muad Ghaith under the Research Incentive Fund (Grant No. 23048). The authors gratefully acknowledge this financial support.